\documentclass[]{JHEP3}
\usepackage{epsfig}

\title{Twisted Moduli and Supersymmetry Breaking}

\author{S. F. King and D. A. J. Rayner \\
Department of Physics and Astronomy,
University of Southampton, Southampton, SO17 1BJ, U.K.}


\abstract{We consider twisted moduli contributions to supersymmetry breaking 
in effective type I string constructions
involving intersecting $D5_i$ and $D9$-branes using Goldstino angles to 
parametrise the supersymmetry breaking.
It is well known that twisted moduli enter at tree-level into
the gauge kinetic functions, and can provide new sources of gaugino
mass if they develop F-term vacuum expectation values.
It is generally assumed that string states
which are sequestered from the twisted moduli receive a zero soft mass
in the twisted modulus domination limit, however the standard
form of K\"{a}hler potential does not reproduce this expectation.
We therefore propose a new form of the 
K\"{a}hler potential which is consistent at leading order with 
the sequestered form proposed by Randall and Sundrum,
and show that it leads to exponentially suppressed sequestered soft masses.
Including the effects of Green-Schwarz mixing, we write down the soft
scalar masses and trilinears arising from a type I string construction
involving  
intersecting $D5_{i}$ and $D9$-branes in the presence of untwisted and twisted
moduli. If the squarks and sleptons are identified with
sequestered states then in the twisted moduli dominated limit
this corresponds to gaugino mediated supersymmetry breaking,
and we discuss two different scenarios for this.
The general results will
be useful for phenomenological studies involving a combination of gravity and 
gaugino mediated SUSY breaking due to the dilaton, untwisted and twisted 
moduli contributions, and enable the soft masses to 
be studied as a function of the different compactification radii.}

\preprint{SHEP 01-29 \\ hep-ph/0111333}

\begin{document}
%
\section{Introduction}

Superstring theories offer the only consistent method for unifying the four
fundamental forces of Nature within a single framework.  Following the
discovery of string dualities~\cite{dual} and
Dirichlet-branes~\cite{dbranes}, heterotic strings
no longer provide the only theory in which to embed the (minimal
supersymmetric) Standard Model (MSSM) and type I (and II) models have
all been considered.  The heterotic and type I models share common
features, but differ in phenomenologically important ways.  For
instance, both scenarios contain the dilaton and (twisted and untwisted)
moduli fields that are related to the geometry of the compactified
space and appear in the low-energy four-dimensional effective SUGRA
theory~\cite{ibanez97}.  However, in type I models the fundamental
string scale $M_{\ast}$ is no longer fixed at the grand unification
$M_{X}$ or Planck scales $M_{P}$, and in principle can be as low as
the TeV scale with the lower bound determined by phenomenology.
An important difference between heterotic and type I is the r\^{o}le played by
twisted moduli fields - closed string states that are {\it trapped} at fixed 
points in the underlying manifold due to the action of orbifold 
compactification. 
Consider the gauge kinetic function $f_{\alpha}$ that appears in the SUGRA 
lagrangian~\cite{ibanez97}.  In weakly coupled heterotic string theory, the
string coupling constant is uniquely determined by the dilaton 
$g_{s}^{-2} \sim Re (S)$, and:
\begin{eqnarray}
 f_{\alpha} = k_{\alpha} S + \Delta_{1-loop}(T_{i})
\end{eqnarray}
where $k_{\alpha}$ is the Kac-Moody level of the gauge factor\footnote{In the
MSSM, $k_{SU(3)_{C}}=k_{SU(2)_{L}}=\frac{3}{5} k_{U(1)_{Y}}=1$.} and
$\Delta_{1-loop}(T_{i})$ arises from 1-loop string threshold 
corrections~\cite{ibnil} ~\footnote{In contrast, there is a very different 
situation for the
strongly coupled case (from M-theory) where $f_{\alpha}$ receives comparable 
contributions at tree-level from the dilaton and untwisted moduli fields
$f \sim S + T$ ~\cite{stronghet}.}.  In 
contrast, type I models have gauge kinetic functions that depend
on the dilaton S for 9-branes {\it and} the moduli fields $T_{i}$ 
for $5_i$-branes, giving rise to different gauge couplings on
different branes. In addition the gauge kinetic functions
have a tree-level dependence on the twisted moduli, and this
gives rise to different gauge couplings even within a particular
D-brane sector.
The tree-level dependence on the twisted moduli fields $Y^{k}$ from 
the $k^{th}$ twisted sector (within the world-volume of a given
D-brane sector) are given by:  
\begin{eqnarray}
 f^{9}_{\alpha} &=& S + \sum_{k} \frac{s_{\alpha, k}}{4 \pi} \sum_{q} Y^{k,q} 
  \label{eq:f9} \\
 f^{5_{i}}_{\beta} &=& T_{i} + \sum_{k} \frac{s_{\beta, k}'}{4 \pi}
   \sum_{p_{i}} Y^{k,p_{i}}  \label{eq:f5i}
\end{eqnarray}
where the gauge coupling is found by the relation:
\begin{eqnarray}
 Re f_{\alpha} = \frac{4 \pi}{g_{\alpha}^{2}}
  \label{eq:gcoupling}
\end{eqnarray}
and $q(p_{i})$ label the fixed points within the $9(5_{i})$-brane and 
$s_{\alpha,k} , s_{\beta,k}'$ are calculable model-dependent
coefficients. 
Thus twisted moduli tend to induce non-universal gauge couplings
even for gauge groups living on a common brane sector.
The twisted moduli also play an important 
r\^{o}le in the cancellation of gauge 
and gravitational anomalies in type I models - like the dilaton in heterotic 
string theory - through a generalised four-dimensional Green-Schwarz 
mechanism~\cite{gs} that mixes twisted and untwisted moduli together.

The K\"{a}hler potential and superpotential can also receive non-perturbative 
contributions, and in the absence of a complete model one may
adopt a phenomenologically-motivated parametrisation in order to 
make progress~\cite{ibanez94,ibanez98}.  The relative contributions to the
overall SUSY breaking F-term vacuum expectation value (vev) from different 
fields can be parametrised in terms of {\it Goldstino} angles. 
In such an approach one
can derive the soft parameters in terms of the Goldstino angles, and
examine various limits in which the dilaton or moduli fields dominate.
As envisaged by the originators of the approach, it may also be used 
to investigate the contributions to SUSY 
breaking from twisted moduli in effective type I theories,
in addition to the usual dilaton and untwisted moduli
fields~\cite{allanach}. However the analyses that have been done so
far have only considered the explicit
situation where the gauge group and matter fields arise from a stack of 
D9-branes, and thus share the same world-volume as all of the twisted moduli 
fields. It is one of the purposes of this paper to 
extend the scope of such analyses to include more general
set-ups involving intersecting $D5_i$ and $D9$-branes. In so doing
we encounter a difficulty that is not present in the case of
a single D9-brane set-up, namely the problem of sequestered
states which do not share the same world volume as the 
twisted moduli, and we show how this problem may be successfully
resolved.

In this paper, then, 
we shall consider twisted moduli contributions to SUSY breaking in
effective type I string constructions based on a general
set-up involving intersecting $D5_i$ and $D9$-branes,
using Goldstino angles to parametrise the SUSY breaking.
It is well known that the F-term vevs
of the twisted moduli fields provide a new source of gaugino masses
~\cite{benakli}. It is also generally assumed that
states that do not live in the
same world-volume should receive zero soft mass contributions
in the twisted moduli dominated limit, which  
offers a possible string realisation of gaugino mediated 
SUSY breaking ($\tilde{g}MSB$)~\cite{benakli},\cite{gaugino}.
However we show that the standard form of K\"{a}hler potential
is not consistent with this physical requirement. 
We therefore propose a new form of the 
K\"{a}hler potential which is consistent at leading order with 
the sequestered form proposed by Randall and Sundrum~\cite{rs},
and which leads to exponentially suppressed sequestered soft masses,
in agreement with physical expectations.
Including the effects of Green-Schwarz mixing we then
write down soft scalar and trilinear masses 
arising from a general string construction involving 
intersecting $D5_i$ and $D9$-branes in the presence of untwisted and twisted
moduli. We show how the results may be applied to $\tilde{g}MSB$
and discuss two explicit scenarios for this. 
The general results will be useful for phenomenological
studies involving a combination of gravity and gaugino mediated 
SUSY breaking due to the dilaton, untwisted and twisted 
moduli contributions, and enable the soft masses
to be studied as a function of the finite compactification radii.

The layout of the remainder of the
paper is as follows. In section \ref{sec:eff}
we discuss effective type I string theories in the presence 
of twisted moduli, point out the difficulty 
with the sequestered soft masses using the standard 
K\"{a}hler potential, and propose a new sequestered form
of K\"{a}hler potential which solves the problem.
In section \ref{sec:anomaly} we generalise our results to include 
Green-Schwarz mixing, then in section \ref{sec:softp} we
write down the resulting soft scalar and trilinear masses
that arise in general string constructions involving
intersecting D5 and D9 branes, in the presence of twisted
and untwisted moduli contributions to SUSY breaking.
In section \ref{sec:simple} we discuss gaugino mediated SUSY
breaking as a simple example, and point out that our results
enable gravity mediated corrections to gaugino mediation to be
studied as a function of the compactification scale.
For completeness we include Appendices on Supergravity basics.

\section{Effective Type I String Theory and Twisted Moduli}  \label{sec:eff}

\subsection{K\"{a}hler Potentials}
In this section we will introduce a generic type I string construction 
involving intersecting $D5_i$-branes 
embedded within $D9$-branes, where coincident 
D-branes give rise to gauge 
groups localised within the world-volume of the corresponding D-brane.  Chiral
charged matter fields appear as open-strings with their ends attached to 
D-branes.  Chan-Paton factors at the string ends carry the gauge quantum 
numbers under the attached gauge group.  This type of construction will lead 
to two distinct types of matter field - $C^{5_{i}}_{j}$ and $C^{9}_{j}$ are 
open strings with both ends attached to the same $D5(9)$-brane, while 
$C^{5_{i} 5_{j}}$ and $C^{9 5_{i}}$ have their ends attached to different 
D-branes and the string tension forces the inverse length of the strings to 
become of order the string scale $M_{\ast}$.  
The $C^{5_{i} 5_{j}}$ states become localised at the 4d 
intersection point between the two D5-branes, while the $C^{9 5_{i}}$ states
have one end attached anywhere along the $5_{i}$-brane world-volume.  The 
spectrum also contains closed strings that correspond to the 
gravity multiplet and dilaton ($S$) and moduli fields ($T_{i}$).  Notice that
this construction is entirely general and is T-dual to alternative 
scenarios involving D7- and D3-branes.  A construction
involving two sets of intersecting branes within a D9-brane is shown in 
figure \ref{fig:const}, but our analysis can be extended for a full set of 
three perpendicular intersecting branes and the open/closed string states 
that result.
\FIGURE[h]{
 \label{fig:const}
 \unitlength=1in
\begin{picture}(4.5,2.7)
\put(-0.2,0){\epsfig{file=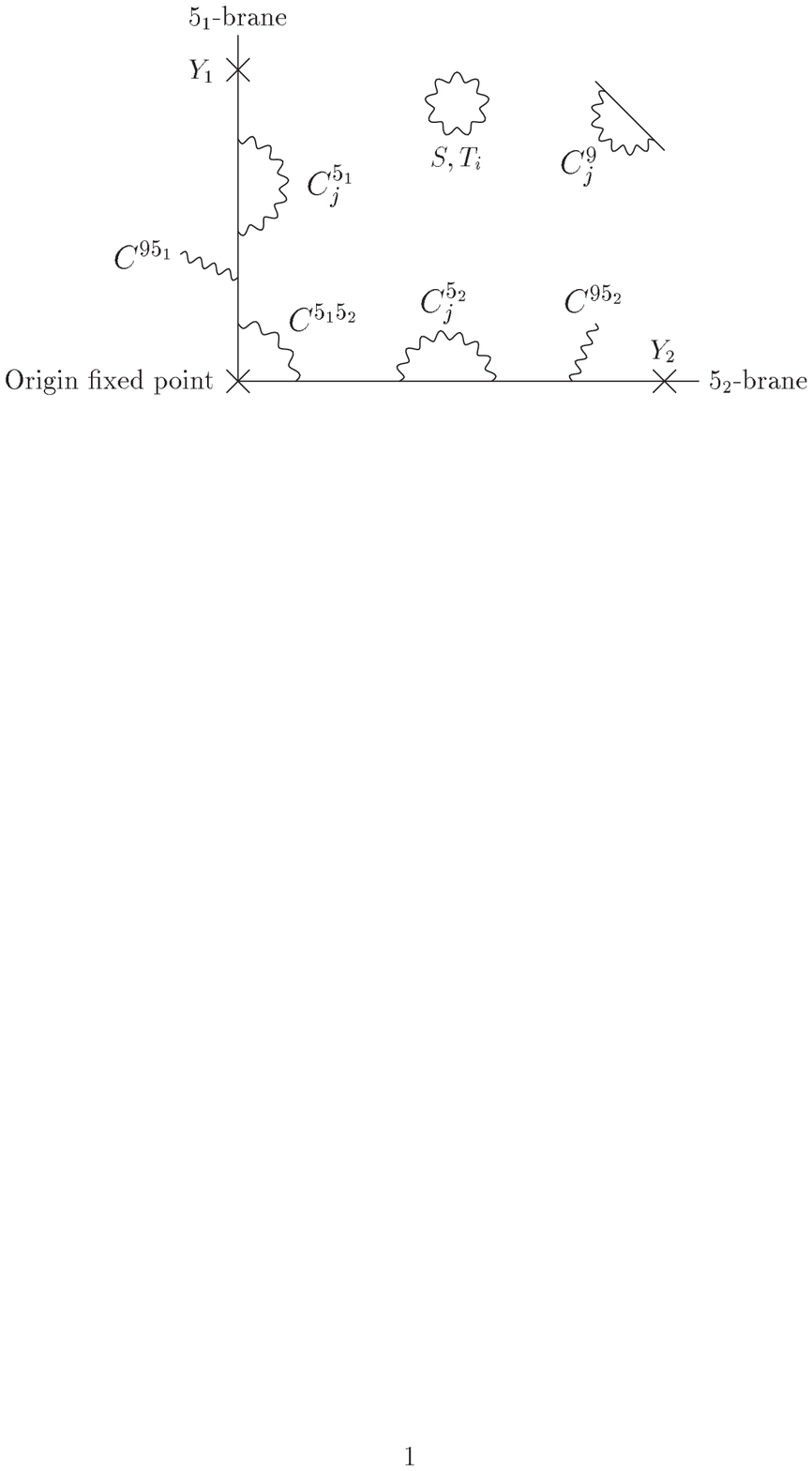, width=5in}}
\end{picture}
   \caption{A generic type I string construction involving two sets of
perpendicular D5-branes embedded within a D9-brane, where the D5-brane 
world-volumes intersect at the origin.  
Charged chiral fields appear as open strings with both ends attached to the 
same D-brane $C^{5_{i}}_{j}$ and $C^{9}_{j}$, or different branes 
$C^{5_{1} 5_{2}}$ and $C^{9 5_{i}}$.  Closed
strings ($S,T_{i}$) can live in the full 10d space, although orbifolding 
leads to closed strings (twisted moduli $Y_{k}$) localised at 4d fixed points 
within the $D5_{i}$-brane world-volume.}
}

We can now exploit the string duality between 10d $SO(32)$ heterotic theory 
and 10d type I theory to derive the 4d K\"{a}hler potential $K(S,T_{i},C_{a})$
for the dilaton, (untwisted) moduli and charged chiral fields that 
arise in the low energy supergravity description of the model with two sets of
intersecting D5-branes embedded within a D9-brane as shown in Figure 
\ref{fig:const}. Ignoring the twisted moduli for the moment, 
the result is~\cite{ibanez98}:
\begin{eqnarray}
 K= -\ln \left( S+\bar{S} - |C^{5_{1}}_{1}|^{2} - |C^{5_{2}}_{2}|^{2} \right)
  -\ln \left( T_{1} + \bar{T_{1}} - |C^{9}_{1}|^{2} 
   - |C^{5_{2}}_{3}|^{2} \right)  \nonumber \hspace{2.5cm} \\
 -\ln \left( T_{2} + \bar{T_{2}} - |C^{9}_{2}|^{2} 
  - |C^{5_{1}}_{3}|^{2} \right) 
   -\ln \left( T_{3} + \bar{T_{3}} - |C^{9}_{3}|^{2} - |C^{5_{1}}_{2}|^{2} 
    - |C^{5_{2}}_{1}|^{2} \right)  \label{eq:kahler1} \hspace*{5mm} \\
 + \frac{ |C^{5_{1} 5_{2}}|^{2} }{ (S+\bar{S})^{1/2} 
  (T_{3}+\bar{T_{3}})^{1/2}} 
  + \frac{ |C^{9 5_{1}}|^{2} }{ (T_{2}+\bar{T}_{2})^{1/2} 
   (T_{3}+\bar{T_{3}})^{1/2}} 
   + \frac{ |C^{9 5_{2}}|^{2} }{ (T_{1}+\bar{T}_{1})^{1/2} 
    (T_{3}+\bar{T_{3}})^{1/2}}  \nonumber 
\end{eqnarray}
The results can easily be extended to include a third $5_{3}$-brane.

Expanding the K\"{a}hler potential in the lowest order in the matter fields\\
\noindent
(i.e. $(S+\bar{S}) \gg |C^{5_{1}}_{1}|^{2} + |C^{5_{2}}_{2}|^{2}$) yields:
\begin{eqnarray}
  \label{eq:kahler2}
 K= -\ln \left( S+\bar{S} \right) -\sum_{i=1}^{3} \ln \left( T_{i} + 
  \bar{T_{i}} \right) + \sum_{i=1}^{2} \frac{|C^{5_{i}}_{i}|^{2}}{(S+\bar{S})}
   + \frac{|C^{5_{2}}_{3}|^{2}}{(T_{1}+\bar{T_{1}})}  
    + \frac{|C^{9}_{1}|^{2}}{(T_{1}+\bar{T_{1}})}  \hspace*{2cm} \\
 + \frac{|C^{5_{1}}_{3}|^{2}}{(T_{2}+\bar{T_{2}})}
  + \frac{|C^{9}_{2}|^{2}}{(T_{2}+\bar{T_{2}})}
   + \frac{|C^{5_{1}}_{2}|^{2}}{(T_{3}+\bar{T_{3}})}
    + \frac{|C^{5_{2}}_{1}|^{2}}{(T_{3}+\bar{T_{3}})}
     + \frac{|C^{9}_{3}|^{2}}{(T_{3}+\bar{T_{3}})}  \hspace*{15mm} \nonumber \\
  + \frac{ |C^{5_{1} 5_{2}}|^{2} }{ (S+\bar{S})^{1/2} 
   (T_{3}+\bar{T_{3}})^{1/2}} 
    + \frac{ |C^{9 5_{1}}|^{2} }{ (T_{2}+\bar{T}_{2})^{1/2} 
     (T_{3}+\bar{T_{3}})^{1/2}} 
      + \frac{ |C^{9 5_{2}}|^{2} }{ (T_{1}+\bar{T}_{1})^{1/2} 
       (T_{3}+\bar{T_{3}})^{1/2}}  \nonumber 
\end{eqnarray}
Using Appendix \ref{sec:sugra} and Eq.\ref{eq:kahexp}, we can identify the 
individual K\"{a}hler metrics (which are diagonal 
$\tilde{K}_{a}=\tilde{K}_{\bar{a}b}\delta_{\bar{a}b}$) for each type of 
charged chiral field:
\begin{eqnarray}
 \tilde{K}_{C^{5_{i}}_{i}} &=& \frac{1}{(S + \bar{S})} \nonumber \\
 \tilde{K}_{C^{5_{i}}_{j}} &=& \frac{1}{(T_{k} + \bar{T_{k}})} \hspace*{3.5cm}
  (i \neq j \neq k \neq i)  \nonumber \\
 \tilde{K}_{C^{9}_{i}} &=& \frac{1}{(T_{i} + \bar{T_{i}})}
  \hspace*{1cm}  \label{eq:kahlerchi} \\
 \tilde{K}_{C^{5_{1} 5_{2}}} &=& \frac{1}{ (S+\bar{S})^{1/2} 
  (T_{3}+\bar{T_{3}})^{1/2}}  \nonumber \\
 \tilde{K}_{C^{9 5_{i}}} &=& \frac{1}{ (T_{j}+\bar{T_{j}})^{1/2}
  (T_{k}+\bar{T_{k}})^{1/2}} \hspace*{1cm} (i \neq j \neq k \neq i)  
   \nonumber 
\end{eqnarray}

The twisted moduli $Y^{k,q}$ 
also contribute to the K\"{a}hler potential, but
the precise form of the contribution is strongly model-dependent.  
For simplicity we shall consider a single twisted modulus 
within each of the three D5-brane sectors, which we denote 
by $Y^k$ where $k=1,2,3$ labels the $D5_k$ branes.
Each of the $Y^k$ may be regarded as a linear combination of
all the twisted moduli within that $D5_k$ brane, so that the
simplified gauge kinetic function is from Eq.\ref{eq:f5i},
\begin{equation}
f^{5_{k}}_{\alpha}=T_{k} + \frac{s_{\alpha}}{4\pi} Y_{k}
\end{equation}
Anomaly cancellation via the Green-Schwarz mechanism suggests that this 
contribution mixes twisted and untwisted moduli together while preserving
modular invariance.  We will work in terms of a general even function 
$\hat{K}$ with an argument:
\begin{equation}
  (Y_{k} + \bar{Y_{k}}) - \delta_{GS}^{k j} \ln (T_{j} + \bar{T}_{j})  
    \label{eq:gs}
\end{equation}
For simplicity, we will initially drop the Green-Schwarz term 
($\delta_{GS}=0$) and assume a very simple form~\cite{simple} 
\begin{equation}
 \hat{K}( Y_{k}, \bar{Y_{k}} ) = \frac{1}{2} (Y_{k} + \bar{Y_{k}})^{2}
   \label{eq:ykahler}
\end{equation}
Hence the tree-level K\"{a}hler potential for the closed string states is:
\begin{eqnarray}
 \bar{K}(h,\bar{h})=-\ln \left( S+\bar{S} \right) 
  - \sum_{i=1}^{3} \ln \left( T_{i} + \bar{T_{i}} \right) 
   + \sum_{k=1}^{3} \hat{K}(Y_{k},\bar{Y}_{k})
\end{eqnarray}
We will repeat our analysis in the presence of a Green-Schwarz mixing term
($\delta_{GS} \neq 0$) in section \ref{sec:anomaly}.

The perturbative superpotential can be expressed in terms of the states 
present in the model by considering the set of renormalisable interactions 
that arise from the splitting and joining of open strings.
\begin{eqnarray}
 W_{ren} = g_{5_{1}} \left( C^{5_{1}}_{1} C^{5_{1}}_{2} C^{5_{1}}_{3}
  + C^{5_{1}}_{3} C^{5_{1} 5_{2}} C^{5_{1} 5_{2}}
   + C^{5_{1}}_{1} C^{9 5_{1}} C^{9 5_{1}} \right) \hspace*{4.5cm} 
  \nonumber \\
 + g_{5_{2}} \left( C^{5_{2}}_{1} C^{5_{2}}_{2} C^{5_{2}}_{3}
  + C^{5_{2}}_{3} C^{5_{1} 5_{2}} C^{5_{1} 5_{2}} 
   + C^{5_{2}}_{2} C^{9 5_{2}} C^{9 5_{2}}\right)
    + g_{5_{3}} C^{5_{1} 5_{2}} C^{9 5_{1}} C^{9 5_{2}}  \hspace*{10mm}
   \label{eq:typeiw} \\
 + g_{9} \left( C^{9}_{1} C^{9}_{2} C^{9}_{3}
  + C^{9}_{1} C^{9 5_{1}} C^{9 5_{1}} + C^{9}_{2} C^{9 5_{2}} C^{9 5_{2}} 
   \right)   \hspace*{5mm}  \nonumber
\end{eqnarray}
where the Yukawa coupling constants (associated with fields arising from each
5-and 9-brane) are given by:
\begin{eqnarray}
 g_{5_{i}}^{2} = \frac{4 \pi}{Re T_{i}} \hspace*{8mm} , \hspace*{8mm}
  g_{9}^{2} = \frac{4 \pi}{Re S}  \label{eq:g5i}
\end{eqnarray}
However, the superpotential can also receive (unknown) non-perturbative
contributions, e.g. from gaugino condensation\footnote{See \cite{abel}
for a recent discussion in the context of stabilising the dilaton potential in
type I string theory.}, that require the F-terms to be parametrised in terms 
of Goldstino angles.

In this general setup, we are assuming that SUSY breaking originates from the 
closed string sector.  In the absence of a Green-Schwarz anomaly cancelling
term in Eq.\ref{eq:ykahler}, the K\"{a}hler metric is diagonal at leading 
order since $Re S, Re T_{i} \gg |C_{a}|^{2}$.  Using 
Eqs.\ref{eq:kahler2},\ref{eq:ykahler},\ref{eq:genf} we can write down the
SUSY breaking F-term vev in terms of two Goldstino angles ($\theta,\phi$), 
where $\theta(\phi)$ describes the relative contributions from the dilaton and
moduli (twisted and untwisted moduli) F-terms respectively, and we are 
assuming a vanishing cosmological constant $V_{0}$.
\begin{eqnarray}
 F_{S} &=& \sqrt{3} m_{3/2} \sin \theta e^{i \alpha_{S}} 
  (K_{\bar{S} S})^{-1/2} = \sqrt{3} m_{3/2} \sin \theta e^{i \alpha_{S}} 
   (S+\bar{S})  \label{eq:ftermsa} \\ 
 F_{T_{i}} &=& \sqrt{3} m_{3/2} \cos \theta \sin \phi \, \Theta_{i} \,
  e^{i \alpha_{i}} (K_{\bar{T_{i}} T_{i}})^{-1/2} = \sqrt{3} m_{3/2} 
   \cos \theta \sin \phi \, \Theta_{i} \, 
    e^{i \alpha_{i}} (T_{i} + \bar{T_{i}})  \hspace*{1cm} 
   \label{eq:ftermsb} \\ 
 F_{Y_{k}} &=& \sqrt{3} m_{3/2} \cos \theta \cos \phi \, \Phi_{k} \, 
e^{i \alpha_{Y_{k}}} (K_{\bar{Y_{k}} Y_{k}})^{-1/2} 
   = \sqrt{3} m_{3/2} \cos \theta \cos \phi \, \Phi_{k} \,
    e^{i \alpha_{Y_{k}}}  \label{eq:ftermsc} 
\end{eqnarray}
where $\sum_{i=1}^{3} \Theta_{i}^{2} =1$ and $\sum_{k=1}^3 \Phi_{k}^{2} =1$.

One can study three limits of phenomenological interest where
different sources of SUSY breaking dominate:
dilaton (S) domination  where $\sin \theta = 1$;
untwisted moduli ($T_{i}$) domination 
where $\cos \theta = \sin \phi = 1$;
and twisted moduli ($Y_{k}$) domination where $\cos \theta = \cos \phi = 1$.
In the next sub-section we shall see that there is a problem with the
sequestered masses in the twisted moduli dominated limit, and then we shall 
show how this problem may be resolved.

\subsection{Problems with the Standard K\"{a}hler Potential}  \label{sec:prob}

In order to illustrate the problem let us consider the case of
a {\it single} linear combination of
twisted moduli located inside the $5_{2}$-brane, which we denote by
$Y_{2}$, corresponding to the simplified gauge kinetic function
$f^{5_{2}}_{\alpha}=T_{2} + s_{\alpha}Y_{2}/4\pi$.
Thus we take the SUSY breaking parameter $\Phi_{2}=1$ in Eq.\ref{eq:ftermsc}.
We regard this linear combination $Y_{2}$ to be
located in the world-volume of the $5_{2}$-brane at
a distance ${\mathcal O}(R_{5_{2}})$ from the intersection states 
$C^{5_{1} 5_{2}}$. Figure \ref{fig:const} shows
that only $C^{5_{2}}_{j} , C^{9}_{j}$ and $C^{9 5_{2}}$ states can couple 
directly to the 
$Y_{2}$ twisted moduli, while $C^{5_{1}}_{j} , C^{9 5_{1}}$
are confined on the $5_{1}$-brane, and $C^{5_{1} 5_{2}}$ is confined
to the origin fixed point. We refer to the states $C^{5_{1}}_{j}, 
C^{9 5_{1}}$ and $C^{5_{1} 5_{2}}$ which are spatially
separated from $Y_{2}$ as being sequestered from it.
Using Eqs.\ref{eq:ftermsa}-\ref{eq:ftermsc},\ref{eq:scmass}
with the standard K\"{a}hler metric for the intersection and $5_{1}$-brane
states of Eq.\ref{eq:kahlerchi}, the sequestered state scalar masses
are found to be 
(still ignoring the Green-Schwarz mixing term $\delta_{GS}=0$):
\begin{eqnarray}
 m_{C^{5_{1} 5_{2}}}^{2} &=&
  m_{3/2}^{2} \left[ 1 - \frac{3}{2}\left( \sin^{2} \theta + \Theta_{3}^{2} \,
   \cos^{2} \theta \, \sin^{2} \phi \right) \right] \nonumber \\
 m_{C^{5_{1}}_{1}}^{2} &=& m_{3/2}^{2} \left[ 1 - 3 \sin^{2} \theta \right] 
  \nonumber \\
 m_{C^{5_{1}}_{2}}^{2} &=& m_{3/2}^{2} \left[ 1 
  - 3 \Theta_{3}^{2} \, \cos^{2}\theta \, \sin^{2}\phi \right] 
   \label{eq:probmass} \\
 m_{C^{5_{1}}_{3}}^{2} &=& m_{3/2}^{2} \left[ 1 
  - 3 \Theta_{2}^{2} \, \cos^{2}\theta \, \sin^{2}\phi \right] 
   \nonumber \\
 m_{C^{9 5_{1}}}^{2} &=& m_{3/2}^{2} \left[ 1 
  - \frac{3}{2} \cos^{2}\theta \, \sin^{2}\phi 
   \left( \Theta_{2}^{2} + \Theta_{3}^{2} \right) \right] \nonumber
\end{eqnarray}
In the twisted moduli dominated limit where
the F-term $F_{Y_{2}}$ is the only 
contribution to the SUSY breaking ($\cos \theta =\cos \phi=1$) 
the intersection state masses from Eq.\ref{eq:probmass} are: 
\begin{eqnarray}
 m_{C^{5_{1} 5_{2}}}^{2} = m_{C^{5_{1}}_{j}}^{2} = m_{C^{9 5_{1}}}^{2}
  = m_{3/2}^{2} \hspace*{1cm} (j=1,2,3)
\label{bad}
\end{eqnarray}
The soft masses in Eq.\ref{bad}
are independent of the separation between the origin and the
fixed point at which the twisted moduli live. This is not what we expect.
Since these states are sequestered from the twisted moduli
we would expect that their soft masses be exponentially 
suppressed by the spatial separation between the two fixed points,
as the following argument explains.

In the twisted moduli dominated limit, the situation 
regarding the sequestered states is physically
equivalent to the gaugino mediated SUSY breaking scenario~\cite{gaugino}
as shown in Figure \ref{fig:gaugino}.
\FIGURE[h]{ 
 \label{fig:gaugino}
 \unitlength=1in
\begin{picture}(6,2)
\put(0,-0.2){\epsfig{file=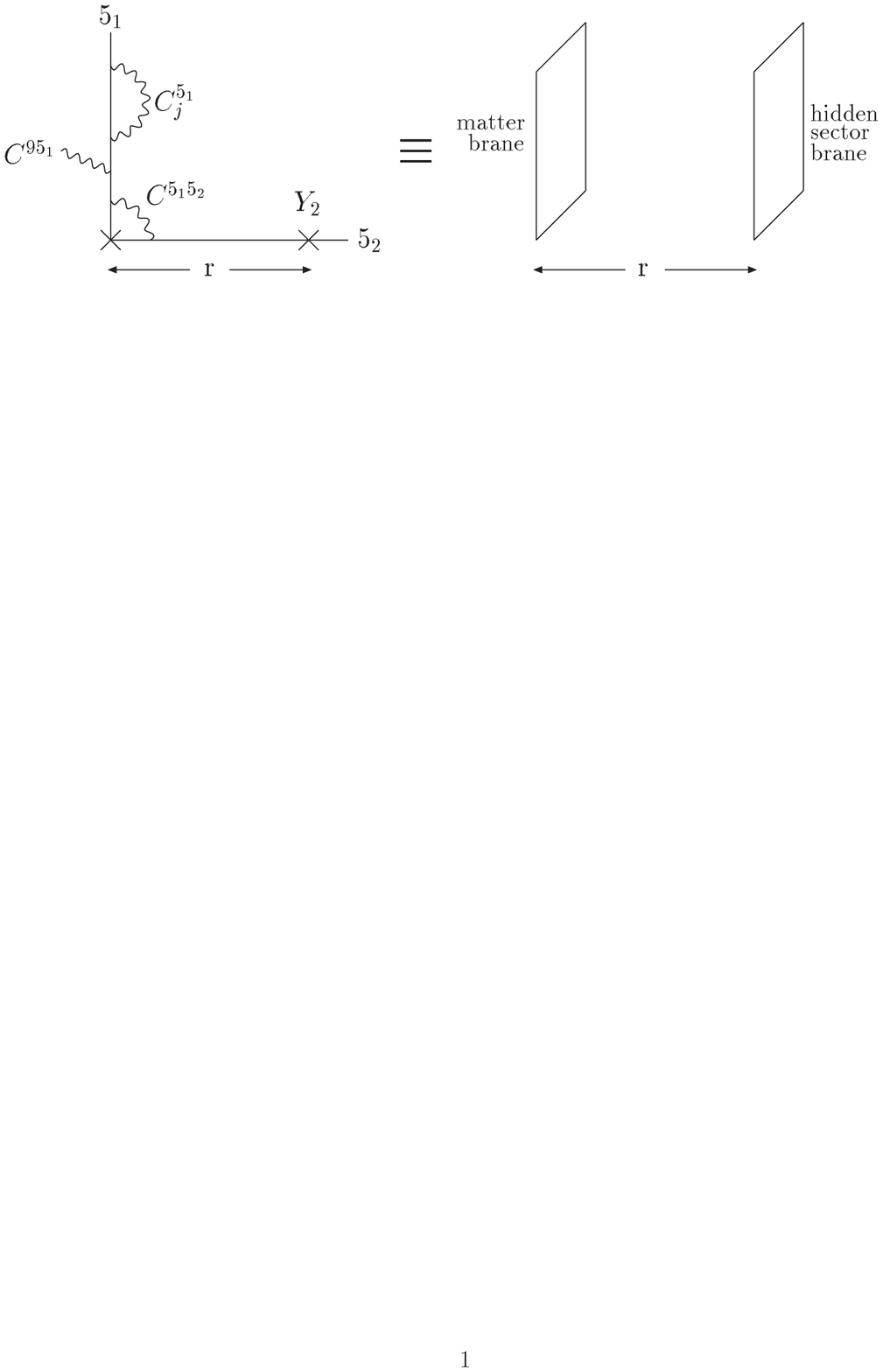, width=6in}}
\end{picture}
 \caption{The intersecting D5-brane construction shares similar
features with the gaugino mediated SUSY breaking model in the limit of a small
compactification radius $R_{5_{1}}$. In this limit, the $C^{5_{1}}_{j}$ and
$C^{9 5_{1}}$ states
are effectively localised at the origin, and these intersection states are
equivalent to the matter brane while the localised twisted modulus
is equivalent to the hidden sector brane where SUSY is broken.  The spatial
separation between the two fixed points (matter and hidden sector brane) is
$r\sim {\mathcal O}(R_{5_{2}})$.}   
}
In gaugino mediation, SUSY is broken on a 4d ``hidden sector 
brane'' which is spatially separated along one (or more) extra dimensions from
another parallel 4d ``matter brane'' where matter fields are
localised. Scalar masses on the matter brane are exponentially 
suppressed at tree-level by the distance between the branes 
but are radiatively generated at one-loop via gaugino
mediation.  In the string theory realisation, the r\^{o}le of the
hidden sector brane is played by the 
twisted moduli localised at a non-trivial fixed point separated from the
origin fixed point which corresponds to the matter brane.

Clearly in the limit of large spatial separation between the two fixed points
$r \sim {\mathcal O}(R_{5_{2}})$ the sequestered soft masses should be 
exponentially suppressed~\footnote{Physically, 
this suppression is due to integrating 
out the heavier modes (with masses above the cutoff $M_{\ast}$) that propagate
between sectors with a Yukawa-like propagator.}:
\begin{equation}
 m_{C^{5_{1} 5_{2}}}^{2} \, , \, m_{C^{5_{1}}_{j}}^{2} \, , \,
  m_{C^{9 5_{1}}}^{2} \sim e^{- M_{\ast} r} m_{3/2}^{2}
\label{exp}
\end{equation}
where $M_{\ast}$ is the ultraviolet cutoff for the effective theory, which we 
will associate with the string scale.  Obviously, a sufficiently large
separation will lead to a negligibly small mass as in $\tilde{g}MSB$ which 
offers a solution to the flavour problems and suppression of flavour-changing
neutral-currents.
In the next sub-section we propose a new form of the
K\"{a}hler potentials which give rise to the correct exponentially
suppressed soft sequestered masses in Eq.\ref{exp} rather than the
result in Eq.\ref{bad}.  In appendix \ref{sec:alt} we consider an
alternative exponential suppression factor
$e^{-\left(M_{\ast}r\right)^{2}}$ that is attributed to
non-perturbative world-sheet instanton corrections~\cite{altexp}.

\subsection{A New K\"{a}hler Potential}  
  \label{sec:sequester1}

We need to modify the intersection state K\"{a}hler potentials
$\tilde{K}_{C^{5_{1} 5_{2}}} , \tilde{K}_{C^{5_{1}}_{j}}$ and 
$\tilde{K}_{C^{9 5_{1}}}$ to give the desired exponentially suppressed mass
prediction in Eq.\ref{eq:probmass} with an explicit dependence on the
separation.  Notice that in the limit of very small separation, we should be 
able to recover the previous (standard) form of Eq.\ref{eq:kahlerchi}.  To 
begin with we will only consider the K\"{a}hler potential for 
$C^{5_{1} 5_{2}}$ states and later generalise to the 
$C^{5_{1}}_{j}, C^{9 5_{1}}$ states as well.

Consider the scalar mass relation of Eq.\ref{eq:scmass} for the intersection
states $C^{5_{1} 5_{2}}$ in the limit of twisted moduli domination
and let us determine the K\"{a}hler potential from the requirement
that it be exponentially suppressed:
\begin{eqnarray}
 m_{C^{5_{1} 5_{2}}}^{2} &=& m_{3/2}^{2} -F_{\bar{Y_{2}}} F_{Y_{2}} 
  \partial_{\bar{Y_{2}}}
  \partial_{Y_{2}} \left( \ln \tilde{K}_{C^{5_{1} 5_{2}}} \right) \nonumber \\
 &=& m_{3/2}^{2} \left[ 1 - \frac{3}{\tilde{K}_{C^{5_{1} 5_{2}}}}
   \partial_{\bar{Y_{2}}} \partial_{Y_{2}} \tilde{K}_{C^{5_{1} 5_{2}}}
  + \frac{3}{(\tilde{K}_{C^{5_{1} 5_{2}}})^{2}}
   \left( \partial_{\bar{Y_{2}}} \tilde{K}_{C^{5_{1} 5_{2}}} \right) 
    \partial_{Y_{2}} \tilde{K}_{C^{5_{1} 5_{2}}} \right]  \hspace*{1cm}
   \label{eq:smass} \\
 &\equiv& e^{- R_{5_{2}} M_{\ast}} m_{3/2}^{2}  \nonumber
\end{eqnarray}
where $F_{Y_{2}} = \sqrt{3} m_{3/2} \cos \theta \cos \phi \, e^{i
\alpha_{Y_{2}}}$; twisted moduli domination corresponds to
$\cos \theta = \cos \phi = 1$,
and $R_{5_{2}}$ is the compactification radius of the second complex dimension
and is of order the separation between the hidden sector $Y_{2}$ moduli and
the intersection states.  The 4d untwisted moduli field $T_{i}$ can be 
decomposed into real and imaginary parts:
\begin{eqnarray}
 T_{i} = \frac{2 R_{i}^{2} M_{\ast}^{2}}{\lambda} + i\eta_{i}
\end{eqnarray}
where $R_{i}^{2}=R_{5_{i}}^{2}$ is the compactification radius on the $i^{th}$
torus, $M_{\ast}$ is the string scale, $\lambda$ is the 10d dilaton which is 
related to the fundamental (perturbative) string coupling which we can set 
equal to unity, and $\eta_{i}$ is an untwisted closed string from the 
Ramond-Ramond sector.  Hence, we can find a relationship between the real part
of the untwisted T-modulus field $T_{2}$ and the compactification radius 
$R_{5_{2}}$:
\begin{eqnarray}
 R_{5_{2}} = \frac{\sqrt{T_{2} + \bar{T}_{2}}}{2 M_{\ast}}
\end{eqnarray} 

We can now solve for the K\"{a}hler potential that
leads to the equivalence of the last two lines of Eq.\ref{eq:smass}:
\begin{eqnarray}
 \tilde{K}_{C^{5_{1} 5_{2}}} = exp \left[ \left( 1-e^{- \sqrt{ T_{2} 
  + \bar{T}_{2}}/2} \right) \frac{(Y_{2}+\bar{Y_{2}})^{2}}{6} \right] \, 
   \zeta[S,T_{1},T_{3}]
\label{solution}
\end{eqnarray}
where $\zeta$ is some arbitrary function of $S,T_{1}$ and/or $T_{3}$.  
The condition that the previous expression for the K\"{a}hler 
potential of Eq.\ref{eq:kahler2} is reproduced
in the limit of a small compactification 
radius, i.e. $R_{5_{2}} \sim \sqrt{T_{2} + \bar{T}_{2}} \longrightarrow 0$
fixes the function $\zeta[S,T_{1},T_{3}]$:
\begin{eqnarray}
 \tilde{K}_{C^{5_{1} 5_{2}}} \longrightarrow \zeta[S,T_{1},T_{3}] \equiv
  \frac{1}{(S+\bar{S})^{1/2} (T_{3}+\bar{T}_{3})^{1/2}}
\end{eqnarray}
Then in the limit of very large 
separation $R_{5_{2}} \sim \sqrt{T_{2}+\bar{T}_{2}} \longrightarrow \infty$:
\begin{eqnarray}
 \tilde{K}_{C^{5_{1} 5_{2}}} \longrightarrow \frac{ e^{(Y_{2}+\bar{Y_{2}})^{2}/6}}{ 
  (S+\bar{S})^{1/2} (T_{3}+\bar{T_{3}})^{1/2}}  \label{eq:kexp1a}
\end{eqnarray}
In this limit,
using Eqs.\ref{eq:ftermsa}-\ref{eq:ftermsc},\ref{eq:scmass} we obtain:
\begin{eqnarray}
 m_{C^{5_{1} 5_{2}}}^{2} =
  m_{3/2}^{2} \left[ 1 - \frac{3}{2}\left( \sin^{2} \theta + \Theta_{3}^{2} \,
   \cos^{2} \theta \, \sin^{2} \phi \right) - \cos^{2}\theta \cos^{2}\phi
    \right]  
\end{eqnarray}
which replaces the form in Eq.\ref{eq:probmass} in the large separation 
limit, and which vanishes in the limit of twisted moduli domination 
$(\cos\theta=\cos\phi=1)$ due to the strong exponential suppression factor. 

In Ref.~\cite{rs}, Randall and Sundrum discussed the conditions under which 
a visible matter sector may be sequestered from a SUSY breaking hidden 
sector and proposed a K\"{a}hler potential that leads to vanishing
scalar masses:
\begin{eqnarray}
 K_{RS} = -3 \tilde{M}_{P}^{2} \ln \left[ 1 + e^{-K_{vis}/3 \tilde{M}_{P}^{2}}
   + e^{-K_{hid}/3 \tilde{M}_{P}^{2}} \right] 
\end{eqnarray}
where $K_{vis}(K_{hid})$ is the separate K\"{a}hler potential for the visible
(hidden) sectors, and $\tilde{M}_{P} = M_{P}/\sqrt{8 \pi}$ is the
reduced Planck mass. If we write the visible sector ($C$) and 
hidden sector ($Y_{2}$) K\"{a}hler potentials as:
\begin{eqnarray}
 K_{vis}= 3 \left| C \right|^{2} \hspace{5mm}, \hspace{5mm}
  K_{hid} = \frac{3}{2} (Y_{2} + \bar{Y_{2}})^{2}
\end{eqnarray}
then the combined K\"{a}hler potential may be expanded 
for small $|C|^2$ and $(Y_2+\bar{Y}_2)^2$ as:
\begin{eqnarray}
 K_{RS} &=& -3 \tilde{M}_{P}^{2} \ln \left[ 1 + e^{- \left| C \right|^{2}/ 
  \tilde{M}_{P}^{2}} + e^{-(Y_{2}+\bar{Y_{2}})^{2}/2\tilde{M}_{P}^{2}} \right]
   \label{eq:kahlerrs2} \\
 &=& \frac{1}{2}(Y_{2}+\bar{Y_{2}})^{2} + {\mathcal O}\left[ 
  (Y_{2}+\bar{Y_{2}})^{4} \right]
   + \left| C \right|^{2} \left[ 1 + \frac{(Y_{2}+\bar{Y_{2}})^{2}}{6 
    \tilde{M}_{P}^{2}} 
     - \frac{(Y_{2}+\bar{Y_{2}})^{4}}{72 \tilde{M}_{P}^{4}} 
      \right] + \ldots  \nonumber
\end{eqnarray}
The expansion of the coefficient of $|C|^2$ in Eq.\ref{eq:kahlerrs2}
is equivalent to the expansion of $e^{(Y_{2}+\bar{Y_{2}})^{2}/6}$
in Eq.\ref{eq:kexp1a} 
up to ${\mathcal O}\left[ (Y_{2}+\bar{Y_{2}})^{4} \right]$,
where we have adopted ``natural'' units and set $\tilde{M}_{P} = 1$.
Therefore the numerator in Eq.\ref{eq:kexp1a} is equivalent to the 
Randall-Sundrum sequestered form of the K\"{a}hler potential 
in Eq.\ref{eq:kahlerrs2} to 
${\mathcal O}\left[ (Y_{2}+\bar{Y_{2}})^{4} \right]$,
which is sufficient for all practical purposes.

We can now write down the modified form of the tree-level K\"{a}hler potential
that yields the ``correct'' mass for the intersection states 
$C^{5_{1} 5_{2}}$ (and similarly the $5_{1}$-brane states $C^{5_{1}}_{j}$ and
$C^{9 5_{1}}$) - in the limit of $Y_{2}$-domination - with an explicit 
dependence on the separation between the intersection point and the twisted 
moduli hidden sector:
\begin{eqnarray}
 K(S,\bar{S},T_{i},\bar{T_{i}},Y_{2},\bar{Y_{2}}) = -\ln \left( S+\bar{S} 
  \right) -\sum_{i=1}^{3} \ln \left( T_{i} + \bar{T_{i}} \right) 
   +\frac{1}{2} (Y_{2} + \bar{Y_{2}})^{2} \hspace*{3cm} \nonumber \\
 + \sum_{C^{5_{1}}_{1}} \frac{\xi(T_{2},Y_{2})}{(S+\bar{S})} |C^{5_{1}}_{1}|^{2}  
  + \sum_{C^{5_{1}}_{2}} \frac{\xi(T_{2},Y_{2})}{(T_{3}
   +\bar{T_{3}})} |C^{5_{1}}_{2}|^{2} 
    + \sum_{C^{5_{1}}_{3}} \frac{\xi(T_{2},Y_{2})}{(T_{2}
     +\bar{T_{2}})} |C^{5_{1}}_{3}|^{2}  \hspace*{2.3cm} 
      \label{eq:kahlermod}  \\
 + \sum_{C^{5_{2}}_{1}} \frac{|C^{5_{2}}_{1}|^{2}}{(T_{3}+\bar{T_{3}})}
  + \sum_{C^{5_{2}}_{2}} \frac{|C^{5_{2}}_{2}|^{2}}{(S+\bar{S})}
   + \sum_{C^{5_{2}}_{3}} \frac{|C^{5_{2}}_{3}|^{2}}{(T_{1}+\bar{T_{1}})} 
  + \sum_{C^{5_{1} 5_{2}}} \frac{\xi(T_{2},Y_{2})}{ 
      (S+\bar{S})^{1/2} (T_{3}+\bar{T_{3}})^{1/2}} |C^{5_{1} 5_{2}}|^{2}
       \nonumber \hspace*{5mm} \\
 + \sum_{C^{9}_{1}} \frac{|C^{9}_{1}|^{2}}{(T_{1}+\bar{T_{1}})}
  + \sum_{C^{9}_{2}} \frac{|C^{9}_{2}|^{2}}{(T_{2}+\bar{T_{2}})}
   + \sum_{C^{9}_{3}} \frac{|C^{9}_{3}|^{2}}{(T_{3}+\bar{T_{3}})}
    \nonumber \hspace*{25mm}  \\
 + \sum_{C^{9 5_{1}}} \frac{\xi(T_{2},Y_{2})}{ 
  (T_{2}+\bar{T_{2}})^{1/2} (T_{3}+\bar{T_{3}})^{1/2}} |C^{9 5_{1}}|^{2}
   + \sum_{C^{9 5_{2}}} \frac{|C^{9 5_{2}}|^{2}}{ 
    (T_{1}+\bar{T_{1}})^{1/2} (T_{3}+\bar{T_{3}})^{1/2}}
       \hspace*{1cm} \nonumber 
\end{eqnarray}
where
\begin{eqnarray}
 \xi(T_{2},Y_{2}) = exp \left[ \left( 1-e^{- \sqrt{ T_{2} 
  + \bar{T}_{2}}/2} \right) \frac{(Y_{2}+\bar{Y_{2}})^{2}}{6} \right] 
\label{xi}
\end{eqnarray}
Before we repeat this analysis with the Green-Schwarz mechanism for anomaly 
cancellation, we will show that our result can be generalised to a 
construction involving three perpendicular D5-branes that all intersect at the
origin.  We will assume that there are three separate linear combinations of
twisted moduli $Y_{i}$ - one combination within each $D5_{i}$-brane 
world-volume - each at a distance ${\mathcal O}(R_{5_{i}})$ from the origin 
intersection states, where
\begin{eqnarray}
 R_{5_{i}} = \frac{\sqrt{T_{i} + \bar{T}_{i}}}{2 M_{\ast}}
\end{eqnarray}
We can immediately write down the form of the K\"{a}hler potential that will 
give the correct prediction for the masses in the Y-dominated SUSY breaking 
limit:
\begin{eqnarray}
 K \supset \frac{\xi(T_{2},Y_{2}) \, \xi(T_{3},Y_{3})}{(S+\bar{S})} 
  |C^{5_{1}}_{1}|^{2} + \frac{\xi(T_{1},Y_{1}) \,
   \xi(T_{3},Y_{3})}{(T_{3} + \bar{T}_{3})} |C^{5_{2}}_{1}|^{2}  
    \hspace*{5cm}  \nonumber \\
  + \frac{\xi(T_{1},Y_{1}) \, \xi(T_{2},Y_{2})}{(T_{2}+\bar{T_{2}})} 
   |C^{5_{3}}_{1}|^{2} +
    \frac{\xi(T_{1},Y_{1}) \, \xi(T_{2},Y_{2}) \, \xi(T_{3},Y_{3}) 
  }{(S+\bar{S})^{1/2} (T_{3}+\bar{T_{3}})^{1/2}} \, |C^{5_{1} 5_{2}}|^{2} 
   \label{eq:kgen}  \hspace*{1cm} \\
 + \frac{\xi(T_{2},Y_{2}) \, \xi(T_{3},Y_{3})}{ 
  (T_{2}+\bar{T_{2}})^{1/2} (T_{3}+\bar{T_{3}})^{1/2}} |C^{9 5_{1}}|^{2}
   + \frac{\xi(T_{1},Y_{1}) \, \xi(T_{3},Y_{3})}{ 
    (T_{1}+\bar{T_{1}})^{1/2} (T_{3}+\bar{T_{3}})^{1/2}}  |C^{9 5_{2}}|^{2}
    + \ldots  \hspace*{1cm} \nonumber 
\end{eqnarray}
Notice that the $C^{5_{i}}_{j}$ and $C^{9 5_{i}}$ states will couple directly 
to the twisted moduli within the same brane ($Y_{i}$), but will receive 
exponentially suppressed SUSY breaking contributions from the twisted moduli 
on different branes ($Y_{k \ne i}$).  The $C^{9}_{j}$ states live in the full
10d space and therefore can couple to all twisted moduli.

We can also generalise our analysis to include more than one twisted moduli
within the world-volumes of each D$5_{i}$-brane that contribute to
SUSY breaking.  For example, consider multiple twisted moduli
$Y_{2}^{a}$ inside the $5_{2}$-brane at a distance $l^{a} \, R_{5_{2}}
\equiv l^{a} \, \sqrt{T_{2}+\bar{T}_{2}}/2 M_{\ast}$ from the origin
fixed point where $C^{5_{1} 5_{2}}$ states are localised.  Hence, the
K\"{a}hler potential for these intersection states includes a sum over
all twisted moduli:
\begin{eqnarray}
 K_{C^{5_{1} 5_{2}}} = \sum_{a} exp \left[ \left( 1-e^{- l^{a} \, \sqrt{ T_{2} 
  + \bar{T}_{2}}/2} \right) \frac{(Y_{2}^{a} +\bar{Y_{2}}^{a})^{2}}{6}
   \right] \, \frac{\left| C^{5_{1} 5_{2}}
    \right|^{2}}{(S+\bar{S})^{1/2} (T_{3}+\bar{T_{3}})^{1/2}}
     \label{solutionmod}
\end{eqnarray}
which we can easily extend for other sequestered states, and
combinations of twisted moduli on different $5_{i}$-branes.

\section{Green-Schwarz Mixing}  \label{sec:anomaly}

In this section we will repeat our previous analysis, but with the inclusion 
of an anomaly cancelling Green-Schwarz term that requires mixing between the
twisted and untwisted moduli fields.  This mixing leads to a non-diagonal
K\"{a}hler metric (at leading order) and we use a canonically normalising
P-matrix in our parametrisation to define SUSY breaking F-terms as discussed 
in section \ref{sec:ft}.

For simplicity, we will again assume that only a single linear combination of 
twisted moduli fields $Y_{2}$ (within the $D5_{2}$-brane world-volume at a 
distance ${\mathcal O}(R_{5_{2}})$ from the intersection point) contributes
to the SUSY breaking.  Since only the twisted moduli from the $5_{2}$-brane
contribute, it is not too unreasonable to suppose that only the $T_{2}$ 
untwisted modulus field participates in the anomaly cancellation.  Using 
Eqs.\ref{eq:gs},\ref{eq:ykahler} we propose that $Y_{2}$ has the following 
K\"{a}hler potential~\cite{simple}:
\begin{equation}
 \hat{K}( Y_{2}, \bar{Y_{2}} ) = \frac{1}{2} \left[ Y_{2}+\bar{Y_{2}} - \delta_{GS} 
  \ln (T_{2} + \bar{T}_{2}) \right]^{2}
   \label{eq:ykahler2}
\end{equation}
We will now calculate the K\"{a}hler metric by using Eq.\ref{eq:kahlermod}
with the modified twisted moduli potential:
\begin{eqnarray}
  K_{\bar{J}I} = \left(
  \begin{array}{ccccc}
   \frac{1}{(S+\bar{S})^{2}} & 0 & 0 & 0 & 0 \\
   0 & \frac{1}{(T_{1}+\bar{T}_{1})^{2}} & 0 & 0 & 0 \\
   0 & 0 & \frac{1}{(T_{2}+\bar{T}_{2})^{2}} \left( k +
    \delta_{GS}^{2} \right) & 0 & - \frac{\delta_{GS}}{T_{2}+\bar{T}_{2}} \\
   0 & 0 & 0 & \frac{1}{(T_{3}+\bar{T}_{3})^{2}} & 0 \\
   0 & 0 & - \frac{\delta_{GS}}{T_{2}+\bar{T}_{2}} & 0 & 1
  \end{array} \right)  \label{eq:kmat}
\end{eqnarray}
where $I=S,T_{i},Y_{2}$ and $k= 1+\delta_{GS} \left[Y_{2}+\bar{Y_{2}} 
 - \delta_{GS} \ln (T_{2} + \bar{T}_{2}) \right]$.  For simplicity, we assume 
that $|C|^{2} \ll (S+\bar{S}), (T_{i}+\bar{T}_{i})$~\footnote{Notice that 
in this limit the exact form of the intersection state K\"{a}hler potential 
$\tilde{K}_{C^{5_{1} 5_{2}}}$ is not important.}.

The K\"{a}hler metric is block-diagonal, and we can canonically normalise the
metric using a $(5 \times 5)$ P-matrix as in ~\cite{allanach}.  Using 
Eq.\ref{eq:genf}, the F-terms are:
\begin{eqnarray}
 F_{I} \equiv \left(
  \begin{array}{c}
   F_{S} \\ F_{T_{1}} \\ F_{T_{2}} \\ F_{T_{3}} \\ F_{Y_{2}} 
  \end{array} \right)
 = \sqrt{3} m_{3/2}  \, P \left(
  \begin{array}{c}
   \sin \theta \, e^{i\alpha_{S}} \\
   \cos \theta \, \sin \phi \, \Theta_{1} \, e^{i\alpha_{1}} \\
   \cos \theta \, \sin \phi \, \Theta_{2} \, e^{i\alpha_{2}} \\
   \cos \theta \, \sin \phi \, \Theta_{3} \, e^{i\alpha_{3}} \\
   \cos \theta \, \cos \phi \, e^{i\alpha_{Y_{2}}} \\
  \end{array} \right)
\label{Fterms}
\end{eqnarray}
where $\theta$ and $\phi$ are Goldstino angles, 
$\sum_{i} \Theta_{i}^{\dagger} \Theta_{i} = 1$ and we have included CP-phases.

Using Eq.\ref{eq:kmat} and imposing the condition 
$P^{\dagger} K_{\bar{J}I} P=1$ we obtain a very complicated expression for the
P-matrix that can be expanded for large values of ($T_{2}+\bar{T}_{2}$) to 
give:
\begin{eqnarray}
  P = \left(
  \begin{array}{ccccc}
   S+\bar{S} & 0 & 0 & 0 & 0 \\
   0 & T_{1}+\bar{T}_{1} & 0 & 0 & 0 \\
   0 & 0 & \frac{T_{2}+\bar{T}_{2}}{\sqrt{k}} & 0 & 
    - \frac{\delta_{GS}}{T_{2}+\bar{T}_{2}} \\
   0 & 0 & 0 & T_{3}+\bar{T}_{3} & 0 \\
   0 & 0 & \frac{\delta_{GS}}{\sqrt{k}} + \frac{\sqrt{k} 
    \delta_{GS}}{(T_{2}+\bar{T}_{2})^{2}} & 0 & 1 
     - \frac{\delta_{GS}^{2}}{(T_{2}+\bar{T}_{2})^{2}}
  \end{array} \right)
 + {\mathcal O} \left[ \frac{1}{(T_{2}+\bar{T}_{2})^{3}} \right]
\label{Pmatrix}
\end{eqnarray}
where $k=1+\delta_{GS} \left[Y_{2}+\bar{Y_{2}} - \delta_{GS} \ln (T_{2} 
+ \bar{T}_{2}) \right]$.
\vspace*{1mm}

From Eqs.\ref{Fterms},\ref{Pmatrix} we find the SUSY breaking F-terms:
\begin{eqnarray}
 F_{S} &=& \sqrt{3} m_{3/2} \sin \theta \, e^{i\alpha_{S}} \, (S+\bar{S}) \nonumber \\
 F_{T_{1}} &=& \sqrt{3} m_{3/2} \cos \theta \sin \phi \, \Theta_{1} 
  \, e^{i\alpha_{1}} \, (T_{1} + \bar{T}_{1}) \nonumber \\
 F_{T_{2}} &=& \sqrt{3} m_{3/2} \cos \theta \left[ \sin \phi \frac{(T_{2}+
  \bar{T}_{2})}{\sqrt{k}} \, \Theta_{2} \, e^{i\alpha_{2}} 
   - \cos \phi \, e^{i\alpha_{Y_{2}}} \, \frac{\delta_{GS}}{T_{2} +\bar{T}_{2}} 
    \right]  \label{eq:fterms3} \\
 F_{T_{3}} &=& \sqrt{3} m_{3/2} \cos \theta \sin \phi \, \Theta_{3} \,
  \, e^{i\alpha_{3}} (T_{3} + \bar{T}_{3}) \nonumber \\
 F_{Y_{2}} &=& \sqrt{3} m_{3/2} \cos \theta \left[ \sin \phi \left( 
  \frac{\delta_{GS}}{\sqrt{k}} + \frac{\sqrt{k} \delta_{GS}}{(T_{2}
   +\bar{T}_{2})^{2}} \right) \Theta_{2} \, e^{i\alpha_{2}} 
    + \cos \phi \, e^{i\alpha_{Y_{2}}} \left( 1 
     - \frac{\delta_{GS}^{2}}{(T_{2} +\bar{T}_{2})^{2}} \right) \right] 
  \nonumber 
\end{eqnarray}
where $F_{T_{2}}$ and $F_{Y_{2}}$ are expanded up to 
${\mathcal O} \left[ \frac{1}{(T_{2}+\bar{T}_{2})^{3}} \right]$.
Notice that in the limit of $T_{2}$ (or $Y_{2}$) modulus domination, 
{\it both} $F_{T_{2}}$ and $F_{Y_{2}}$ are non-zero.  
Setting $\cos \theta = \cos \phi =1$ still corresponds to the
$Y_{2}$-domination limit, even in the presence of Green-Schwarz
mixing, and we expect the intersection state masses
to depend on the separation from the $Y_{2}$-fields as before.  

Our previous analysis, in the absence of a Green-Schwarz mixing term, 
leads us to propose the following generalisation of the 
K\"{a}hler potential $\tilde{K}_{C^{5_{1} 5_{2}}}$
in Eq.\ref{solution}:
\begin{eqnarray}
 \tilde{K}_{C^{5_{1} 5_{2}}} = \frac{ exp \left[ \frac{1}{6} \left( 
  1-e^{- \sqrt{ T_{2} + \bar{T}_{2}}/2} \right) 
   \left\{ Y_{2}+\bar{Y_{2}} - \delta_{GS} \ln (T_{2} + \bar{T}_{2}) 
    \right\}^{2} \right] }{(S+\bar{S})^{1/2} (T_{3}+\bar{T_{3}})^{1/2}}
     \label{eq:kint}
\end{eqnarray}
which leads to an exponentially suppressed intersection state mass in
the limit of $Y_{2}$-domination:
\begin{eqnarray}
 m_{C^{5_{1} 5_{2}}}^{2} = e^{- \sqrt{T_{2}+\bar{T}_{2}}/2} \, \, m_{3/2}^{2}
  + {\mathcal O} \left[ \frac{1}{(T_{2}+\bar{T}_{2})^{3/2}} \right]
\end{eqnarray}
Similar results apply to the $C^{5_{1}}_{j}$ and $C^{9 5_{1}}$ states 
and we obtain a K\"{a}hler potential as in
Eq.\ref{eq:kahlermod}, but with Eq.\ref{xi} generalised to
\begin{eqnarray}
 \xi(T_{2},Y_{2}) = exp \left[ \frac{1}{6} \left( 
  1-e^{- \sqrt{ T_{2} + \bar{T}_{2}}/2} \right) 
   \left\{ Y_{2}+\bar{Y_{2}} - \delta_{GS} \ln (T_{2} + \bar{T}_{2}) 
    \right\}^{2} \right]  \label{eq:xi2}
\end{eqnarray}
In section \ref{sec:simple} we will consider an explicit example and analyse 
the soft parameters in various limits.

Notice that our comment at the end of section \ref{sec:sequester1} about
including the effects of multiple SUSY breaking twisted moduli still holds.  
The previous expression of Eq.\ref{eq:kgen} is easily generalised by 
replacing the arguments as follows:
\begin{eqnarray} 
 (Y_{k}+\bar{Y}_{k}) \longrightarrow (Y_{k}+\bar{Y}_{k})
  - \delta_{GS} \ln (T_{k} + \bar{T}_{k}).
\end{eqnarray} 

\section{Soft SUSY Breaking Parameters}  \label{sec:softp}

We will now write down the complete list of soft scalar masses and trilinears
that arise in a general string construction involving two intersecting
D5-branes embedded within a D9-brane, where a single linear combination of 
twisted moduli $Y_2$ is located at
a fixed point within the world-volume of one of the branes ($5_{2}$).  
It is straightforward to generalise these results to
more twisted moduli fields, and we have explicitly discussed
the case of three twisted moduli $Y_i$ in section \ref{sec:eff}.
The results presented in this section will be useful for performing
more general phenomenological analyses of sparticle spectra
in string theory than have been done so far in the
literature~\footnote{In appendix \ref{sec:alt} we consider an
alternative exponential suppression factor
$e^{-\left(M_{\ast}r\right)^{2}}$ that can be attributed to
non-perturbative world-sheet instanton corrections~\cite{altexp}, and
we calculate modified expressions for scalar masses and trilinears.}.

Note that
the gaugino masses require knowledge of the gauge group embedding, and 
therefore gaugino masses are more model-dependent.  We will consider a simple
example in section \ref{sec:simple}.

\subsection{Scalar Masses}  \label{sec:gsm}

Using Eqs.\ref{eq:kahlermod},\ref{eq:fterms3},\ref{eq:xi2},\ref{eq:scmass}
we can write down the scalar masses for the 
non-sequestered states $C^{5_{2}}_{j} , C^{9}_{j}$ and
$C^{9 5_{2}}$ which 
couple directly to the twisted moduli $Y_{2}$:
\begin{eqnarray}
 m_{C^{5_{2}}_{1}}^{2} = m_{C^{9}_{3}}^{2}
  &=& m_{3/2}^{2} \left[ 1 - 3 \Theta_{3}^{2} \,
   \cos^{2} \theta \, \sin^{2} \phi \right] \nonumber \\
 m_{C^{5_{2}}_{2}}^{2} &=& m_{3/2}^{2} \left[ 1 - 3 \sin^{2} \theta \right] 
   \nonumber \\
 m_{C^{5_{2}}_{3}}^{2} = m_{C^{9}_{1}}^{2} 
  &=& m_{3/2}^{2} \left[ 1 - 3 \Theta_{1}^{2} \,
   \cos^{2} \theta \, \sin^{2} \phi \right]  \label{eq:52mass} \\
 m_{C^{9}_{2}}^{2} &=& m_{3/2}^{2} \left[ 1 - 3 \Theta_{2}^{2} \,
   \cos^{2} \theta \, \sin^{2} \phi \right] \nonumber \\
 m_{C^{9 5_{2}}}^{2} &=& m_{3/2}^{2} \left[ 1 
  - \frac{3}{2} \cos^{2}\theta \, \sin^{2}\phi 
   \left( \Theta_{1}^{2} + \Theta_{3}^{2} \right) \right] \nonumber
\end{eqnarray}

\noindent The sequestered states 
$C^{5_{2}}_{j} , C^{5_{1}5_{2}}$ and $C^{9 5_{1}}$ are 
spatially separated from the twisted modulus field $Y_2$ and have
masses of the form:
\begin{eqnarray} 
 m_{C^{5_{1} 5_{2}}}^{2} &=& \tilde{m}^{2} - \frac{3}{2} m_{3/2}^{2} 
  \left( \sin^{2} \theta + \Theta_{3}^{2} \, \cos^{2} \theta \, \sin^{2} \phi 
   \right)  \nonumber \\
 m_{C^{9 5_{1}}}^{2} &=& \tilde{m}^{2} - \frac{3}{2} m_{3/2}^{2} \,
  \cos^{2}\theta \, \sin^{2}\phi \left( \frac{\Theta_{2}^{2}}{k}
   + \Theta_{3}^{2} \right) \nonumber \\
m_{C^{5_{1}}_{1}}^{2} &=& \tilde{m}^{2} - 3 m_{3/2}^{2} \sin^{2}\theta 
   \label{eq:51mass} \\
 m_{C^{5_{1}}_{2}}^{2} &=& \tilde{m}^{2} - 3 m_{3/2}^{2} \Theta_{3}^{2} 
  \cos^{2}\theta \sin^{2}\phi  \nonumber \\
 m_{C^{5_{1}}_{3}}^{2} &=& \tilde{m}^{2} - \frac{3}{k} m_{3/2}^{2} 
  \Theta_{2}^{2} \cos^{2}\theta \sin^{2}\phi  \nonumber
\end{eqnarray}
where 
\begin{eqnarray}
 \tilde{m}^{2} = m_{3/2}^{2} \left[ \vrule width 0pt height 18pt
  1 - \cos^{2}\theta \cos^{2}\phi
    \left( 1-e^{-\sqrt{T_{2}+\bar{T}_{2}}/2} \right) \right. \hspace*{7cm} 
     \nonumber \\
 - \frac{\cos^{2}\theta \sin^{2}\phi \, \Theta_{2}^{2} \, \delta_{GS}}{k}
  \left( 1- e^{-\sqrt{T_{2}+\bar{T}_{2}}/2} \right)
   \left\{ Y_{2}+\bar{Y_{2}}-\delta_{GS} \ln(T_{2}+\bar{T}_{2}) \right\} 
    \hspace*{1.2cm}  \label{mtilde} \\
 + \frac{\cos^{2}\theta \sin^{2}\phi \, \Theta_{2}^{2} \, 
  e^{-\sqrt{T_{2}+\bar{T}_{2}}/2}
  }{32 k} \sqrt{T_{2}+\bar{T}_{2}} \left\{ Y_{2}+\bar{Y_{2}}-\delta_{GS} 
   \ln(T_{2}+\bar{T}_{2}) \right\}^{2} \, \left( 2 +
    \sqrt{T_{2}+\bar{T}_{2}} \right) \hspace*{0.8cm} \nonumber \\
 - \frac{\cos^{2}\theta \cos\phi \sin\phi \, \left( \Theta_{2} \, 
  e^{i(\alpha_{2}-\alpha_{Y_{2}})} + \Theta_{2}^{\dagger} \, 
   e^{-i(\alpha_{2}-\alpha_{Y_{2}})} \right) e^{-\sqrt{T_{2}+\bar{T}_{2}}/2} 
    }{32 \sqrt{k} \, (T_{2}+\bar{T}_{2})} 
     \left\{ Y_{2}+\bar{Y_{2}}-\delta_{GS} \ln(T_{2}+\bar{T}_{2}) \right\} 
      \hspace*{-0.4cm} \nonumber \\
 \left. \times \left( \vrule width 0pt height 15pt
  8 (T_{2}+\bar{T}_{2})^{3/2} + \delta_{GS} 
   \left\{ Y_{2}+\bar{Y_{2}}-\delta_{GS} \ln(T_{2}+\bar{T}_{2}) \right\} 
    \vrule width 0pt height 16pt\right)
     \vrule width 0pt height 18pt \right]  
 + {\mathcal O}\left[ \frac{\delta_{GS}}{(T_{2}+\bar{T}_{2})^{3/2}} \right]
  \hspace*{2cm} \nonumber
\end{eqnarray}
In the limit of a large separation, 
$R_{5_{2}} \sim \sqrt{T_{2}+\bar{T}_{2}} \rightarrow \infty$ 
\begin{eqnarray}
 \tilde{m}^{2} \rightarrow m_{3/2}^{2} \left[ \vrule width 0pt height 18pt
  1 - \cos^{2}\theta \cos^{2}\phi - \frac{\cos^{2}\theta \sin^{2}\phi \, 
  \Theta_{2}^{2} \, \delta_{GS}}{k} \left\{ Y_{2}+\bar{Y_{2}}-\delta_{GS} 
   \ln(T_{2}+\bar{T}_{2}) \right\} \right] 
\end{eqnarray}
and for a small separation $R_{5_{2}} \sim \sqrt{T_{2}+\bar{T}_{2}} 
\rightarrow 0$ 
\begin{eqnarray}
 \tilde{m}^{2} \rightarrow m_{3/2}^{2}
\end{eqnarray}

Now we will consider the different limits of SUSY breaking:

\noindent $\bullet$ Dilaton domination $(\sin\theta=1)$:
\begin{eqnarray}
 m_{C^{5_{2}}_{1 , 3}}^{2} &=& m_{C^{5_{1}}_{2 , 3}}^{2} =
  m_{C^{9}_{1 , 2 , 3}}^{2} = m_{C^{9 5_{1 , 2}}}^{2} = m_{3/2}^{2} 
   \nonumber \\
 m_{C^{5_{2}}_{2}}^{2} &=& m_{C^{5_{1}}_{1}}^{2} = -2m_{3/2}^{2}  
  \label{eq:scmassdil} \\
 m_{C^{5_{1} 5_{2}}}^{2} &=& -\frac{1}{2} m_{3/2}^{2}  \nonumber
\end{eqnarray}
Notice that this limit generally gives rise to tachyonic states.

\noindent $\bullet$ T-moduli domination $(\cos\theta=\sin\phi=1)$:
\begin{eqnarray}
 m_{C^{5_{2}}_{1}}^{2} = m_{C^{9}_{3}}^{2}
  &=& m_{3/2}^{2} \left( 1 - 3 \Theta_{3}^{2} \right) 
   \hspace*{5mm} , \hspace*{5mm}  
 m_{C^{5_{2}}_{2}}^{2} = m_{3/2}^{2} \hspace*{5mm} , \hspace*{5mm}
 m_{C^{5_{2}}_{3}}^{2} = m_{C^{9}_{1}}^{2} 
  = m_{3/2}^{2} \left( 1 - 3 \Theta_{1}^{2} \right)  \nonumber \\
 m_{C^{9}_{2}}^{2}
  &=& m_{3/2}^{2} \left( 1 - 3 \Theta_{2}^{2} \right) 
   \hspace*{5mm} , \hspace*{5mm} 
 m_{C^{5_{1} 5_{2}}}^{2} = \tilde{m}_{T}^{2} - \frac{3}{2} m_{3/2}^{2}
  \Theta_{3}^{2}  \label{eq:scmasstmod} \\
 m_{C^{9 5_{2}}}^{2} &=& m_{3/2}^{2} \left[ 1 - \frac{3}{2} 
  \left( \Theta_{1}^{2} + \Theta_{3}^{2} \right) \right] \hspace*{5mm} , 
   \hspace*{5mm} 
 m_{C^{9 5_{1}}}^{2} =  \tilde{m}_{T}^{2} - \frac{3}{2} m_{3/2}^{2}
  \left( \frac{\Theta_{2}^{2}}{k} + \Theta_{3}^{2} \right)  \nonumber \\
 m_{C^{5_{1}}_{1}}^{2} &=& \tilde{m}_{T}^{2}  \hspace*{5mm} , \hspace*{5mm}
 m_{C^{5_{1}}_{2}}^{2} = \tilde{m}_{T}^{2} - 3 m_{3/2}^{2} \Theta_{3}^{2} 
  \hspace*{5mm} , \hspace*{5mm}
 m_{C^{5_{1}}_{3}}^{2} = \tilde{m}_{T}^{2} - \frac{3}{k} m_{3/2}^{2} 
  \Theta_{2}^{2} \nonumber
\end{eqnarray}
where
\begin{eqnarray}
 \tilde{m}_{T}^{2} = m_{3/2}^{2} \left[ \vrule width 0pt height 18pt
  1 - \frac{\Theta_{2}^{2} \, \delta_{GS}}{k}
  \left( 1- e^{-\sqrt{T_{2}+\bar{T}_{2}}/2} \right)
   \left\{ Y_{2}+\bar{Y_{2}}-\delta_{GS} \ln(T_{2}+\bar{T}_{2}) \right\} 
   \right. \nonumber  \hspace*{4mm} \\
 \left. + \frac{\Theta_{2}^{2} \, e^{-\sqrt{T_{2}+\bar{T}_{2}}/2}
  }{32 k} \sqrt{T_{2}+\bar{T}_{2}} \left\{ Y_{2}+\bar{Y_{2}}-\delta_{GS} 
   \ln(T_{2}+\bar{T}_{2}) \right\}^{2} \, \left( 2 +
    \sqrt{T_{2}+\bar{T}_{2}} \right) \right]  \nonumber
\end{eqnarray}

\noindent $\bullet$ $Y_{2}$-moduli domination $(\cos\theta=\cos\phi=1)$:
\begin{eqnarray}
 m_{C^{5_{2}}_{1 , 2 , 3}}^{2} &=& m_{C^{9}_{1 , 2 , 3}}^{2} 
  = m_{C^{9 5_{2}}}^{2} = m_{3/2}^{2}  \nonumber \\
 m_{C^{5_{1} 5_{2}}}^{2} &=& m_{C^{5_{1}}_{1 , 2 , 3}}^{2} =
  m_{C^{9 5_{1}}}^{2} = e^{-\sqrt{T_{2}+\bar{T}_{2}}/2} \, m_{3/2}^{2} 
   \label{eq:scmassymod} 
\end{eqnarray}
where the $5_{2}$-and 9-brane states couple directly to the SUSY breaking 
twisted moduli and are not exponentially suppressed.

Physically the twisted $Y_2$ moduli dominated limit, corresponds
to gaugino mediated SUSY breaking, if the standard model
quark and lepton states are identified with the sequestered
states (see later example). The dilaton and T-moduli domination
limits correspond to different examples of gravity mediated
SUSY breaking. In the general case where one is not
in any particular limit, SUSY breaking will have
contributions from the F-terms of the dilaton and untwisted moduli 
as well as the twisted moduli, and then one must use the general
formulae for the scalar masses in Eqs.\ref{eq:52mass}, \ref{eq:51mass}.

\subsection{Trilinears}  \label{sec:gtri}

The trilinear and Yukawa couplings arise from the superpotential, where the
dominant tree-level contributions are shown in Eq.\ref{eq:typeiw} in terms 
of open string states.  The precise structure of the Yukawa and trilinear 
matrices depend on the identification of these string states with MSSM fields.
The leading terms are constrained by string selection rules and gauge 
invariance.  Higher order corrections can be generated by higher-dimensional 
operators where powers of the model cutoff (e.g. the string or Planck scales) 
lead to a large suppression.  We will illustrate how different identifications
lead to alternative Yukawa structures in section \ref{sec:simple}, as recently
discussed in Ref.~\cite{gordy}.

We will now list the dominant trilinear couplings that arise from the 
perturbative superpotential of Eq.\ref{eq:typeiw}. Using 
Eqs.\ref{eq:kahlermod},\ref{eq:fterms3},\ref{eq:xi2},\ref{eq:trilinear2} 
and making the standard assumption that the Yukawa 
couplings $Y_{abc}$ have no dependence on the dilaton and 
moduli fields ($\partial_{I} \ln Y_{abc}=0$)~\footnote{Notice that by 
definition, the Yukawa couplings are related to the moduli fields through 
Eq.\ref{eq:g5i} so this assumption is not really valid, but we 
make it for illustrative purposes so that our results may be compared to 
others in the literature.} we find:
\begin{eqnarray}
 A_{C^{5_{2}}_{3} C^{5_{1}5_{2}} C^{5_{1}5_{2}}} 
  = -\sqrt{3} m_{3/2} \, \cos\theta \left[ \sin\phi \,
  \frac{\Theta_{2} \, e^{i\alpha_{2}}}{\sqrt{k}}  \right. \hspace*{6.6cm}
   \nonumber \\
 + \sin\phi \, \frac{\Theta_{2} \, e^{i\alpha_{2}}}{12\sqrt{k}} \,
  e^{-\sqrt{T_{2}+\bar{T}_{2}}/2} \, \sqrt{T_{2}+\bar{T}_{2}}
   \left\{ Y_{2}+\bar{Y_{2}}-\delta_{GS} 
   \ln(T_{2}+\bar{T}_{2}) \right\}^{2}  \label{eq:aii23} \hspace*{1.5cm} \\
 \left. - \cos\phi \, \frac{e^{i\alpha_{Y_{2}}}}{3}
  \left( 1+ 2 \, e^{-\sqrt{T_{2}+\bar{T}_{2}}/2} \right) 
   \left\{ Y_{2}+\bar{Y_{2}}-\delta_{GS} \ln(T_{2}+\bar{T}_{2}) \right\} 
\right] \hspace*{1cm}
  \nonumber \\
 A_{C^{9}_{1} C^{9 5_{1}} C^{9 5_{1}}} 
  = -\sqrt{3} m_{3/2} \, \left[ \sin\theta \, e^{i\alpha_{S}}
   \right. \hspace*{8.7cm} \nonumber \\
 + \cos\theta \sin\phi \, \frac{\Theta_{2} \, e^{i\alpha_{2}}}{12\sqrt{k}} \,
  e^{-\sqrt{T_{2}+\bar{T}_{2}}/2} \, \sqrt{T_{2}+\bar{T}_{2}}
   \left\{ Y_{2}+\bar{Y_{2}}-\delta_{GS} 
   \ln(T_{2}+\bar{T}_{2}) \right\}^{2}  \label{eq:a9951951} \hspace*{1.5cm} \\
 \left. - \cos\theta \cos\phi \, \frac{e^{i\alpha_{Y_{2}}}}{3}
  \left( 1+ 2 \, e^{-\sqrt{T_{2}+\bar{T}_{2}}/2} \right) 
   \left\{ Y_{2}+\bar{Y_{2}}-\delta_{GS} \ln(T_{2}+\bar{T}_{2}) \right\} 
    \right] \hspace*{1cm} \nonumber \\
 A_{C^{5_{1} 5_{2}} C^{9 5_{1}} C^{9 5_{2}}} 
  = -\sqrt{3} m_{3/2} \, \left[ \frac{1}{2} \sin\theta \, e^{i\alpha_{S}}
   + \frac{1}{2} \cos\theta\sin\phi \left( \Theta_{1} \, e^{i\alpha_{1}} +
  \frac{\Theta_{2} \, e^{i\alpha_{2}}}{\sqrt{k}} 
   - \Theta_{3} \, e^{i\alpha_{3}} \right) \right. \hspace*{2mm} \nonumber \\
 + \cos\theta \sin\phi \, \frac{\Theta_{2} \, e^{i\alpha_{2}}}{12\sqrt{k}} \,
  e^{-\sqrt{T_{2}+\bar{T}_{2}}/2} \, \sqrt{T_{2}+\bar{T}_{2}}
   \left\{ Y_{2}+\bar{Y_{2}}-\delta_{GS} 
   \ln(T_{2}+\bar{T}_{2}) \right\}^{2}  \label{eq:ai951952} \hspace*{1cm} \\
 \left. - \cos\theta \cos\phi \, \frac{e^{i\alpha_{Y_{2}}}}{3}
  \left( 1+ 2 \, e^{-\sqrt{T_{2}+\bar{T}_{2}}/2} \right) 
   \left\{ Y_{2}+\bar{Y_{2}}-\delta_{GS} \ln(T_{2}+\bar{T}_{2}) \right\} 
    \right] \hspace*{5mm} \nonumber \\
 A_{C^{5_{1}}_{3} C^{5_{1}5_{2}} C^{5_{1}5_{2}}} = 
  A_{C^{5_{1}}_{1} C^{5_{1}}_{2} C^{5_{1}}_{3}} = 
   A_{C^{5_{1}}_{1} C^{9 5_{1}} C^{9 5_{1}}} 
    = -\sqrt{3} m_{3/2} \, \cos\theta \left[ \sin\phi \, \Theta_{1} \, 
     e^{i\alpha_{1}} \right. \nonumber \hspace*{1.8cm} \\
 + \sin\phi \frac{\Theta_{2} \, e^{i\alpha_{2}}}{8\sqrt{k}} \, 
  e^{-\sqrt{T_{2}+\bar{T}_{2}}/2} \, \sqrt{T_{2}+\bar{T}_{2}}
   \left\{ Y_{2}+\bar{Y_{2}}-\delta_{GS} 
    \ln(T_{2}+\bar{T}_{2}) \right\}^{2}  \label{eq:aii13} \hspace*{1.5cm} \\
 \left. - \cos\phi \, e^{i\alpha_{Y_{2}}}
  \, e^{-\sqrt{T_{2}+\bar{T}_{2}}/2} 
   \left\{ Y_{2}+\bar{Y_{2}}-\delta_{GS} \ln(T_{2}+\bar{T}_{2}) \right\} 
    \right] \hspace*{1cm} \nonumber \\
 A_{C^{5_{2}}_{1} C^{5_{2}}_{2} C^{5_{2}}_{3}} 
  = A_{C^{5_{2}}_{2} C^{9 5_{2}} C^{9 5_{2}}} 
   = -\sqrt{3} m_{3/2} \, \cos\theta \left[ \sin\phi \,
   \frac{\Theta_{2} \, e^{i\alpha_{2}}}{\sqrt{k}} \right. \hspace{4.6cm} 
    \nonumber \\
 \left. - \cos\phi \, e^{i\alpha_{Y_{2}}}
  \left\{ Y_{2}+\bar{Y_{2}}-\delta_{GS} \ln(T_{2}+\bar{T}_{2}) \right\} 
   \right] \label{eq:a212223} \hspace*{1.5cm} \\
 A_{C^{9}_{1} C^{9}_{2} C^{9}_{3}} = A_{C^{9}_{2} C^{9 5_{2}} C^{9 5_{2}}} 
  = -\sqrt{3} m_{3/2} \, \left[ \sin\theta \, e^{i\alpha_{S}} \right.
   \hspace*{6.7cm} \nonumber \\
 \left. - \cos\theta\cos\phi \, e^{i\alpha_{Y_{2}}}
  \left\{ Y_{2}+\bar{Y_{2}}-\delta_{GS} \ln(T_{2}+\bar{T}_{2}) \right\} 
   \right] \label{eq:a9123} \hspace*{1.5cm}
\end{eqnarray}

Now we will consider the different limits of SUSY breaking:

\noindent $\bullet$ Dilaton domination $(\sin\theta=1)$:
\begin{eqnarray}
 A_{C^{9}_{1} C^{9 5_{1}} C^{9 5_{1}}} 
  = A_{C^{9}_{1} C^{9}_{2} C^{9}_{3}} 
   = A_{C^{9}_{2} C^{9 5_{2}} C^{9 5_{2}}} 
    = 2 \, A_{C^{5_{1}5_{2}} C^{9 5_{1}} C^{9 5_{2}}}
     = -\sqrt{3} m_{3/2} \, e^{i\alpha_{S}}  \nonumber \\
 A_{C^{5_{2}}_{3} C^{5_{1}5_{2}} C^{5_{1}5_{2}}} 
  = A_{C^{5_{1}}_{3} C^{5_{1}5_{2}} C^{5_{1}5_{2}}} 
   = A_{C^{5_{1}}_{1} C^{5_{1}}_{2} C^{5_{1}}_{3}} =0  \hspace*{2.6cm}
    \label{eq:asmod} \\
 A_{C^{5_{1}}_{1} C^{9 5_{1}} C^{9 5_{1}}} =
  A_{C^{5_{2}}_{2} C^{9 5_{2}} C^{9 5_{2}}} 
   = A_{C^{5_{2}}_{1} C^{5_{2}}_{2} C^{5_{2}}_{3}} = 0  \hspace*{2.6cm}
    \nonumber
\end{eqnarray}

\noindent $\bullet$ T-moduli domination $(\cos\theta=\sin\phi=1)$:
\begin{eqnarray}
 A_{C^{5_{2}}_{3} C^{5_{1}5_{2}} C^{5_{1}5_{2}}} 
 &=& -\sqrt{\frac{3}{k}} \, m_{3/2} \, \Theta_{2} \, e^{i\alpha_{2}} -\hat{A} 
   \nonumber \\
 A_{C^{9}_{1} C^{9 5_{1}} C^{9 5_{1}}} &=& -\hat{A} \nonumber \\
 A_{C^{5_{1} 5_{2}} C^{9 5_{1}} C^{9 5_{2}}} 
  &=& -\frac{\sqrt{3}}{2} m_{3/2} \left( \Theta_{1} \, e^{i\alpha_{1}} +
  \frac{\Theta_{2} \, e^{i\alpha_{2}}}{\sqrt{k}} 
   - \Theta_{3} \, e^{i\alpha_{3}} \right) - \hat{A} \hspace*{1cm}
  \nonumber \\
 A_{C^{5_{1}}_{3} C^{5_{1}5_{2}} C^{5_{1}5_{2}}} &=& 
  A_{C^{5_{1}}_{1} C^{5_{1}}_{2} C^{5_{1}}_{3}} 
   = A_{C^{5_{1}}_{1} C^{9 5_{1}} C^{9 5_{1}}}
   = -\sqrt{3} m_{3/2} \, \Theta_{1} \, e^{i\alpha_{1}} 
    - \frac{3}{2} \hat{A}  \label{eq:atmod} \\
 A_{C^{5_{2}}_{1} C^{5_{2}}_{2} C^{5_{2}}_{3}} 
  &=& A_{C^{5_{2}}_{2} C^{9 5_{2}} C^{9 5_{2}}} 
    = -\sqrt{\frac{3}{k}} \, m_{3/2} \, \, \Theta_{2} \, e^{i\alpha_{2}}
     \nonumber \\
 A_{C^{9}_{1} C^{9}_{2} C^{9}_{3}} &=& A_{C^{9}_{2} C^{9 5_{2}} C^{9 5_{2}}} 
  = 0 \nonumber
\end{eqnarray}
where
\begin{eqnarray}
 \hat{A} = m_{3/2} \frac{\Theta_{2} \, e^{i\alpha_{2}}}{4\sqrt{3k}} \, 
  e^{-\sqrt{T_{2}+\bar{T}_{2}}/2} \, \sqrt{T_{2}+\bar{T}_{2}}
   \left\{ Y_{2}+\bar{Y_{2}}-\delta_{GS} 
    \ln(T_{2}+\bar{T}_{2}) \right\}^{2}  
\end{eqnarray}

\noindent $\bullet$ $Y_{2}$-moduli domination $(\cos\theta=\cos\phi=1)$:
\begin{eqnarray}
 A_{C^{5_{2}}_{3} C^{5_{1}5_{2}} C^{5_{1}5_{2}}} 
  &=& A_{C^{9}_{1} C^{9 5_{1}} C^{9 5_{1}}} =
   A_{C^{5_{1} 5_{2}} C^{9 5_{1}} C^{9 5_{2}}} =
    \left( 1+ 2 \, e^{-\sqrt{T_{2}+\bar{T}_{2}}/2} \right) \tilde{A}
     \nonumber \\
 A_{C^{5_{1}}_{3} C^{5_{1}5_{2}} C^{5_{1}5_{2}}} 
  &=& A_{C^{5_{1}}_{1} C^{5_{1}}_{2} C^{5_{1}}_{3}} 
   = A_{C^{5_{1}}_{1} C^{9 5_{1}} C^{9 5_{1}}} 
   = 3 e^{-\sqrt{T_{2}+\bar{T}_{2}}/2} \, \tilde{A}  \label{eq:aymod} \\
 A_{C^{5_{2}}_{1} C^{5_{2}}_{2} C^{5_{2}}_{3}} 
  &=& A_{C^{5_{2}}_{2} C^{9 5_{2}} C^{9 5_{2}}} 
   =  A_{C^{9}_{1} C^{9}_{2} C^{9}_{3}} 
    = A_{C^{9}_{2} C^{9 5_{2}} C^{9 5_{2}}} = 3 \tilde{A} \nonumber 
\end{eqnarray}
where
\begin{eqnarray}
 \tilde{A} = m_{3/2} \, \frac{e^{i\alpha_{Y_{2}}}}{\sqrt{3}}
  \left\{ Y_{2}+\bar{Y_{2}}-\delta_{GS} \ln(T_{2}+\bar{T}_{2}) \right\} 
   \label{eq:ahat}
\end{eqnarray}

\section{Gaugino Mediated SUSY Breaking}  
  \label{sec:simple}

The preceeding results are based on a general set-up of intersecting $D5_i$
and $D9$ branes. In order to discuss $\tilde{g}MSB$ it is sufficient
to specialize to the case of just two
intersecting sets of $D5$ branes, $5_1$ and $5_2$.
This set-up arises for example in the explicit string constructions
of \cite{shiutye}. We shall assume that the
MSSM gauge group arises from the $5_{2}$-brane only.
This enables approximate gauge coupling unification
to be achieved.
The MSSM matter fields are identified as either 
$C^{5_{1}5_{2}}$ or $C^{5_{2}}_{j}$ states.
We assume that any $C^{5_{1}5_{2}}$ states
are gauge singlets with respect to any gauge groups on the
$5_{1}$-brane. Such a set-up may be achieved in practice by constructions 
involving severely asymmetric compactifications (for example $R_{5_{2}} \gg 
R_{5_{1}}$~\cite{rayner}), where the combined gauge groups generally arise 
from linear combinations of groups on each set of branes.  The asymmetry 
ensures that the dominant contributions live on the $5_{2}$-brane,
a limit we refer to as ``single brane dominance''.

Although the perpendicular $5_{1}$-brane seems to be irrelevant
in this scenario, in fact it plays an important r\^{o}le since
the $C^{5_{1}5_{2}}$ states are sequestered
at a distance $r \sim {\mathcal O}(R_{5_{2}})$ 
from the fixed point associated with the twisted modulus 
$Y_{2}$. From Eq.\ref{eq:51mass} we find the soft mass
for the sequestered state to be
\begin{eqnarray} 
 m_{C^{5_{1} 5_{2}}}^{2} &=& \tilde{m}^{2} - \frac{3}{2} m_{3/2}^{2} 
  \left( \sin^{2} \theta + \Theta_{3}^{2} \, \cos^{2} \theta \, \sin^{2} \phi 
   \right)  \label{eq:5152exmass}
\end{eqnarray}
where $\tilde{m}^{2}$ is given in Eq.\ref{mtilde}.
In the twisted moduli dominated limit $(\cos\theta=\cos\phi=1)$
\begin{equation}
 m_{C^{5_{1} 5_{2}}}^{2} = 
  e^{-\sqrt{T_{2}+\bar{T}_{2}}/2} \, m_{3/2}^{2}.
\end{equation}
If the standard model states are all identified as intersection states
$C^{5_{1}5_{2}}$ then for large radius of compactification this
corresponds to gaugino mediated SUSY breaking.
However the soft mass in Eq.\ref{eq:5152exmass} is valid away from the
twisted modulus dominated limit, and also is valid for a
small compactification radius. It therefore allows more general
and detailed studies of gaugino mediation to be performed,
including the effects of finite radius of compactification, and
the contributions from gravity mediation effects, which in type I
theories correspond to the dilaton and untwisted moduli F-term vevs.

The non-sequestered soft masses are given by
Eq.\ref{eq:52mass},
\begin{eqnarray}
 m_{C^{5_{2}}_{1}}^{2} = m_{C^{9}_{3}}^{2}
  &=& m_{3/2}^{2} \left[ 1 - 3 \Theta_{3}^{2} \,
   \cos^{2} \theta \, \sin^{2} \phi \right]  \nonumber \\
 m_{C^{5_{2}}_{2}}^{2} &=& m_{3/2}^{2} \left[ 1 - 3 \sin^{2} \theta \right] 
  \label{eq:52exmass} \\
 m_{C^{5_{2}}_{3}}^{2} 
  &=& m_{3/2}^{2} \left[ 1 - 3 \Theta_{1}^{2} \,
   \cos^{2} \theta \, \sin^{2} \phi \right]   \nonumber
\end{eqnarray}

The MSSM gauge groups all arise 
from the $5_{2}$-brane, and using Eq.\ref{eq:f5i} with a single linear 
combination of twisted moduli fields within the $5_{2}$-brane world-volume, 
we find:
\begin{eqnarray}
 f_{\alpha}^{5_{2}}=T_{2} + \frac{s_{\alpha}}{4\pi} Y_{2} \hspace*{2cm} 
  \left( \alpha=SU(3)_{C} , SU(2)_{L} , U(1)_{Y} \right)
   \label{eq:fsm}
\end{eqnarray}
where $s_{\alpha}$ are model-dependent coefficients that depend on the details
of the orbifold compactification.  Notice that for $Z_{3}$ and $Z_{7}$ 
orbifolds, these coefficents are proportional to the MSSM 1-loop beta-function
coefficients $b_{\alpha}$.

We can find the gaugino masses using
Eqs.\ref{eq:gcoupling},\ref{eq:fterms3},\ref{eq:gauginomass2}:
\begin{eqnarray}
 M_{\alpha} = \frac{\sqrt{3} m_{3/2} \, g_{\alpha}^{2}}{8\pi} \cos \theta
  \left[ \sin \phi \, \Theta_{2} \, e^{i \alpha_{2}} 
   \left\{ \frac{T_{2} + \bar{T_{2}}}{\sqrt{k}}
 + \frac{s_{\alpha}}{4\pi} \left( \frac{\delta_{GS}}{\sqrt{k}} 
  + \frac{\sqrt{k} \, \delta_{GS}}{(T_{2}+\bar{T}_{2})^{2}} \right) \right\}
   \right. \hspace*{0.5cm} \\
 \left. - \cos \phi \, e^{i\alpha_{Y_{2}}} 
  \left\{ \frac{\delta_{GS}}{T_{2}+\bar{T}_{2}} -
   \frac{s_{\alpha}}{4\pi} \left( 1-\frac{\delta_{GS}^{2}}{(T_{2}+
    \bar{T}_{2})^{2}}  \right) \right\} \right]  \hspace*{0cm} \nonumber 
\end{eqnarray}

Now consider different limits of SUSY breaking:

\noindent $\bullet$ Dilaton domination $(\sin\theta=1)$:
\begin{eqnarray}
 M_{\alpha} = 0 \hspace*{2cm} 
  \left( \alpha=SU(3)_{C} , SU(2)_{L} , U(1)_{Y} \right)
\end{eqnarray}
\noindent $\bullet$ T-modulus domination $(\cos\theta=\sin\phi=1)$:
\begin{eqnarray}
 M_{\alpha} = \frac{\sqrt{3} m_{3/2} \, g_{\alpha}^{2}}{8\pi}
  \Theta_{2} \, e^{i \alpha_{2}} 
   \left\{ \frac{T_{2} + \bar{T_{2}}}{\sqrt{k}}
 + \frac{s_{\alpha}}{4\pi} \left( \frac{\delta_{GS}}{\sqrt{k}} 
  + \frac{\sqrt{k} \, \delta_{GS}}{(T_{2}+\bar{T}_{2})^{2}} \right) \right\}
\end{eqnarray}
\noindent $\bullet$ $Y_{2}$-modulus domination $(\cos\theta=\cos\phi=1)$:
\begin{eqnarray}
 M_{\alpha} = -\frac{\sqrt{3} m_{3/2} \, g_{\alpha}^{2}}{8\pi} e^{i\alpha_{Y_{2}}}
  \left\{ \frac{\delta_{GS}}{T_{2}+\bar{T}_{2}} -
   \frac{s_{\alpha}}{4\pi} \left( 1-\frac{\delta_{GS}^{2}}{(T_{2}+
    \bar{T}_{2})^{2}}  \right) \right\}
\end{eqnarray}

\subsection{Scenario A - Gaugino Mediated SUSY Breaking
For All Three Families}  \label{sec:gmsb}

\FIGURE[h]{
 \label{fig:modela}
 \unitlength=1in
\begin{picture}(4,2.4)
\put(0.4,0){\epsfig{file=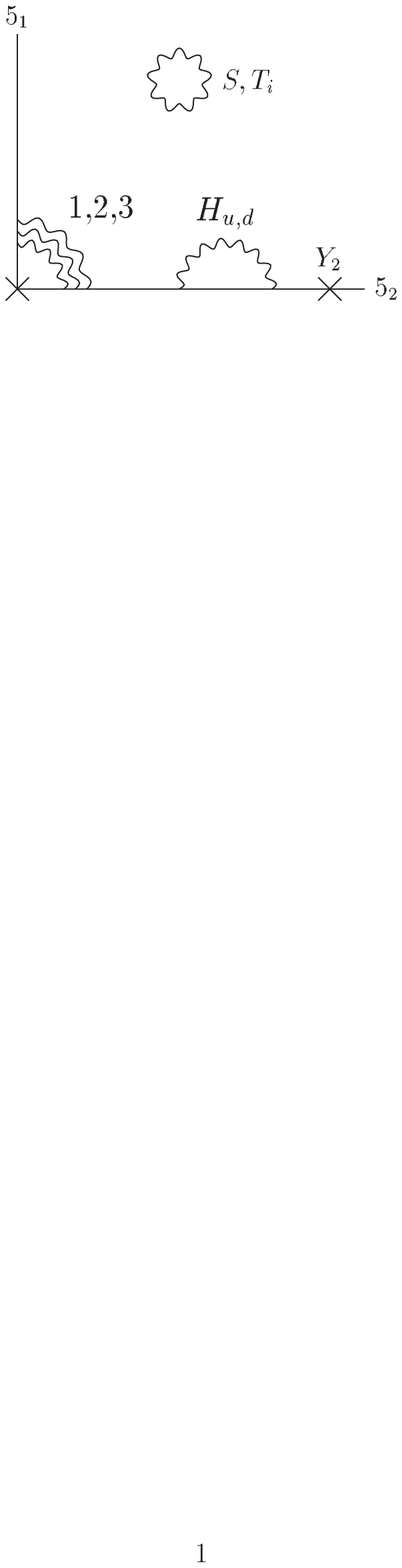, width=3in}}
\end{picture}
   \caption{The allocation of charged chiral fields in scenario A 
which is similar to the gaugino mediated SUSY breaking model~\cite{gaugino}.
The MSSM gauge group arises from the $5_{2}$-brane, and all three MSSM chiral 
families are localised at the origin, while the Higgs and MSSM gauge fields 
live on the $5_{2}$-brane.  The dilaton and moduli fields
$S,T_{i}$ live in the full 10d space and a single twisted moduli $Y_{2}$ is 
localised at a fixed point inside the $5_{2}$-brane world-volume.}
}

In scenario A, depicted in Figure \ref{fig:modela},
the three chiral families are open string states localised at 
the origin fixed point, and the Higgs fields feel two extra dimensions as open
string states with both ends attached to the $5_{2}$-brane.
\begin{eqnarray}
 Q_{i} \, , \, L_{i} \, , \, U_{i} \, , \, D_{i} \, , \, E_{i} 
  &\equiv& C^{5_{1}5_{2}} \hspace*{1cm} (i=1,2,3)  \label{eq:all}\\
 H_{u} \, , \, H_{d} &\equiv& C^{5_{2}}_{j}  \nonumber 
\end{eqnarray}
The Higgs states carry an extra index that plays an important r\^{o}le in 
constructing the perturbative superpotential from open string states.  The 
tree-level superpotential of Eq.\ref{eq:typeiw} contains the terms:
\begin{eqnarray}
 W_{ren} =  {\mathcal O}(g_{5_{2}}) \left[ 
  C^{5_{2}}_{1} C^{5_{2}}_{2} C^{5_{2}}_{3} 
   + C^{5_{2}}_{3} C^{5_{1}5_{2}} C^{5_{1}5_{2}} \right]
\end{eqnarray}
In order to obtain non-zero third family Yukawa couplings (at tree-level), we 
can immediately see that the Higgs fields must be $C^{5_{2}}_{3}$ states. This
leads to a ``democratic'' Yukawa texture (and trilinear matrix) where all 
entries are equal:
\begin{eqnarray}
 Y_{ij}^{ude} \sim {\mathcal O}(g_{5_{2}}) \left(
  \begin{array}{ccc}
   1 & 1 & 1 \\
   1 & 1 & 1 \\
   1 & 1 & 1 
  \end{array} \right)
\end{eqnarray}
The democratic structure arises due to the presence of three 
(indistinguishable) chiral families, localised at the origin fixed point 
($C^{5_{1}5_{2}}$).  However, type I compactifications do not generally lead 
to low-energy spectra with this property as one (or more) families generally 
arise with both ends attached to the same D5-brane ($C^{5_{i}}$) which is the
situation in scenario B.

The squark and slepton and higgs soft masses are given by
Eqs.\ref{eq:5152exmass}, \ref{eq:52exmass} with the identifications
in Eq.\ref{eq:all}. In general the squark and slepton ($C^{5_{1}5_{2}}$)
soft masses receive unsuppressed
contributions from the dilaton and untwisted moduli F-term vevs,
which corresponds to the string version of normal gravity mediation.
In the limit of twisted moduli domination, we
see that the quarks and lepton states acquire exponentially small soft scalar 
masses:
\begin{eqnarray}
 m_{\tilde{Q}_{i}}^{2} = m_{\tilde{U}_{i}}^{2} = m_{\tilde{D}_{i}}^{2} = 
  m_{\tilde{L}_{i}}^{2} = m_{\tilde{E}_{i}}^{2} 
   = e^{-\sqrt{T_{2}+\bar{T}_{2}}/2} \, m_{3/2}^{2} \hspace*{1cm} (i=1,2,3)
\end{eqnarray}
and the Higgs scalars obtain much larger masses due to their direct coupling 
with the SUSY breaking sector (twisted moduli):
\begin{eqnarray}
 m_{h_{u}}^{2} = m_{h_{d}}^{2} = m_{3/2}^{2} 
\end{eqnarray}
This yields the same spectrum as the gaugino mediated SUSY breaking 
scenario~\cite{gaugino}, where vanishingly small scalar masses (due to the 
separation between sectors) offers an attractive (and natural) solution to 
the SUSY flavour problem~\footnote{Flavour-changing neutral-current 
suppression places a lower limit on the size of the separation.}. However, 
unlike $\tilde{g}MSB$, the third family trilinear 
$A_{33} \equiv A_{C^{5_{2}}_{3} C^{5_{1}5_{2}} C^{5_{1}5_{2}}}$ from 
Eq.\ref{eq:aymod} is not loop suppressed and depends on the explicit
function of twisted moduli $\hat{K}(Y_{2},\bar{Y_{2}},T_{2},\bar{T}_{2})$.

The general results in Eqs.\ref{eq:5152exmass}, \ref{eq:52exmass}
enable us to smoothly move away from the twisted moduli
dominated limit (corresponding to gaugino mediated SUSY breaking)
and consider the contributions of the dilaton and untwisted moduli
to the soft masses (corresponding to the gravity contributions
to SUSY breaking). We can also consider the effect of smoothly
changing the compactification radius $R_{5_{2}}$
(corresponding to varying the distance $r$ in Figure \ref{fig:gaugino}.)

Notice that if we assign the Higgs fields as different $5_{2}$-brane states - 
for example $H_{u} \equiv C^{5_{2}}_{3}$ and $H_{d} \equiv C^{5_{2}}_{1,2}$ -
then it is possible to generate a $\mu$-term in the tree-level 
superpotential if we add a gauge singlet that acquires a non-zero vev 
($N \equiv C^{5_{2}}_{1}$ say).
\begin{eqnarray}
 W_{ren} \supset N H_{u} H_{d} \longrightarrow \mu \sim \langle N \rangle
\end{eqnarray}

\subsection{Scenario B - Gaugino Mediated SUSY Breaking
For the First and Second Families Only}  \label{sec:bmsb}

In scenario B, depicted in Figure \ref{fig:modelb},
the third family is moved on to the $5_{2}$-brane along with 
the Higgs and gauge fields.
\begin{eqnarray}
 Q_{i} \, , \, L_{i} \, , \, U_{i} \, , \, D_{i} \, , \, E_{i} 
  &\equiv& C^{5_{1}5_{2}} \hspace*{1cm} (i=1,2)  \label{eq:allb}\\
   Q_{3} \, , \, L_{3} \, , \, U_{3} \, , \, D_{3} \, , \, E_{3} \, , \, 
    H_{u} \, , \, H_{d} &\equiv& C^{5_{2}}_{j}  \nonumber 
\end{eqnarray}
The separation of the third family from the first two chiral families appears
frequently in type I string compactifications.  Ref.~\cite{rayner} provides 
the motivation for this scenario in the limit of a vanishing $5_{1}$-brane
compactification radius.  In this case, the gauge groups are dominated by 
their components on the $5_{2}$-brane.  Notice that an extended Pati-Salam
gauge group appears instead of the MSSM group, and gauge invariance with 
respect to this larger symmetry prevents proton decay (and first and second
family Yukawa couplings at tree-level) by forbidding R-parity violating 
operators.

In order to generate a third family Yukawa coupling at tree-level, the Higgs 
and third family singlets and doublets must carry different indices.  However,
we are still free to choose whether the Higgs fields allocation can give rise 
to first and second family Yukawa couplings at tree-level.  For example,
suppose that we choose the following allocations:
\begin{eqnarray}
 Q_{3} \, , \, L_{3} \equiv C^{5_{2}}_{1} \hspace*{1cm} 
  U_{3} \, , \, D_{3} \, , \, E_{3} \equiv C^{5_{2}}_{2} \hspace*{1cm} 
   H_{u} \, , \, H_{d} \equiv C^{5_{2}}_{3}  
\end{eqnarray}
\FIGURE[h]{
 \label{fig:modelb}
 \unitlength=1in
\begin{picture}(4,2.4)
\put(0.4,0){\epsfig{file=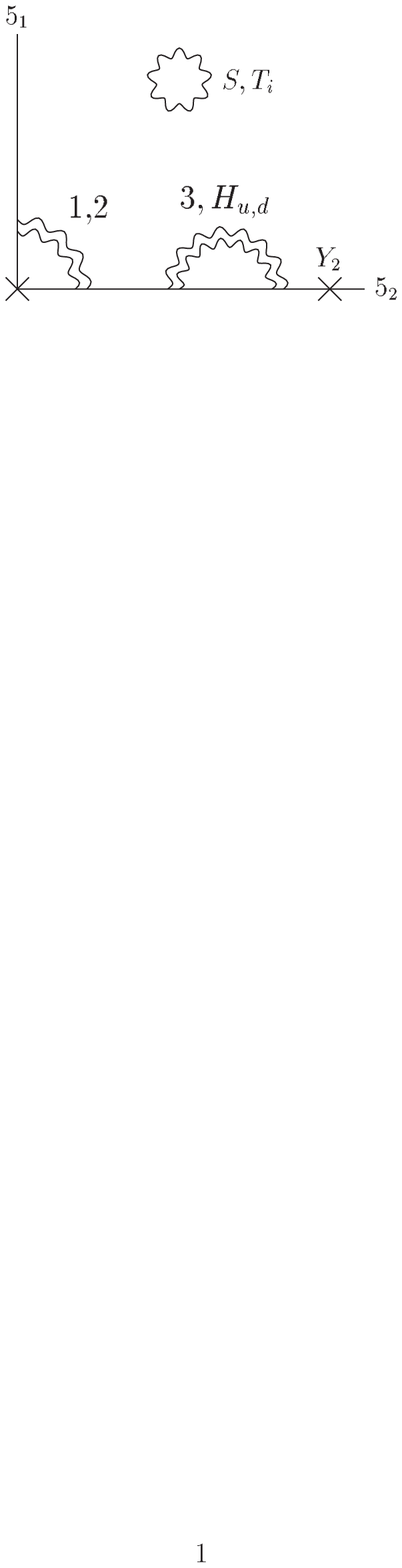, width=3in}}
\end{picture}
   \caption{The allocation of charged chiral fields in scenario B 
which is similar to the brane mediated SUSY breaking model~\cite{rayner}.  
The MSSM gauge group arises from the $5_{2}$-brane, and only the first two 
chiral families live at the origin while the third family, Higgs and gauge 
fields live on the $5_{2}$-brane. The dilaton and moduli fields $S,T_{i}$ live
in the full 10d space and a single twisted moduli $Y_{2}$ is 
localised at a fixed point inside the $5_{2}$-brane world-volume.}
}

We will generate block-diagonal Yukawa textures that are not consistent with
experimental data:
\begin{eqnarray}
 Y_{ij}^{ude} \sim {\mathcal O}(g_{5_{2}}) \left(
  \begin{array}{ccc}
   1 & 1 & 0 \\
   1 & 1 & 0 \\
   0 & 0 & 1 
  \end{array} \right)
\end{eqnarray}

However, if we choose that $H_{u,d} \neq C^{5_{2}}_{3}$, then we generate a
Yukawa texture with only a single non-zero value in the (33) entry:
\begin{eqnarray}
 Y_{ij}^{ude} \sim {\mathcal O}(g_{5_{2}}) \left(
  \begin{array}{ccc}
   0 & 0 & 0 \\
   0 & 0 & 0 \\
   0 & 0 & 1 
  \end{array} \right)
\end{eqnarray}
that is more compatible with data, as higher order corrections can generate
the required structure.  We also obtain a trilinear matrix with a single (33)
entry:
\begin{eqnarray}
  \tilde{A}_{ij}^{ude} =A_{ij} Y_{ij}^{ude} \sim \left(
  \begin{array}{ccc}
   0 & 0 & 0 \\
   0 & 0 & 0 \\
   0 & 0 & A_{33} 
  \end{array} \right)
\end{eqnarray}
where $A_{33}=A_{C^{5_{2}}_{1} C^{5_{2}}_{2} C^{5_{2}}_{3}}$. 

The soft masses for this scenario are very similar to scenario A,
except that the third family is now a $C^{5_{2}}_{2}$ state,
so is now a non-sequestered state.
In the twisted moduli domination limit of Eq.\ref{eq:scmassymod}, the first 
two families receive exponentially suppressed masses as for scenario A.  
However, the third family and Higgs acquire large soft masses 
$\sim {\mathcal O}(m_{3/2})$ which may be read off from
Eq.\ref{eq:52exmass}.

Notice that experimental constraints from 
flavour-changing neutral-current data is only sensitive to the first two 
families, and this scenario (with a hierarchically larger third family) may
not violate these constraints, thereby providing an interesting
alternative solution to the flavour-changing problem.

\section{Conclusions}  \label{sec:conc}

We have considered twisted moduli contributions to supersymmetry breaking in
effective type I string constructions based on
intersecting $D5_i$ and $D9$-branes, using the formalism
of Goldstino angles and extending the scope 
of previous analyses which were based on a single $D9$-brane sector.
The more general set-up allows the possibility of states
which are sequestered from twisted moduli states which are
located at fixed points and cannot move freely.
The sequestered states should have suppressed soft mass contributions
from distant twisted moduli, and this observation has been used
to suggest how $\tilde{g}MSB$ might be implemented in type I string
theory \cite{benakli}. However, contrary to this expectation, we found 
that the standard form of the K\"{a}hler 
potential leads to non-zero soft masses for the sequestered
states in the twisted moduli dominated limit. This motivated
us to look for a new form of K\"{a}hler potential for the
sequestered states. We have proposed a new form of the 
K\"{a}hler potential which is consistent at leading order with 
the sequestered form proposed by Randall and Sundrum~\cite{rs},
and which leads to exponentially suppressed sequestered soft masses.
Including the effect of Green-Schwarz mixing
we have written down soft scalar and trilinear masses 
arising from a general string construction involving 
intersecting $D5_i$ and $D9$-branes in the presence of untwisted and twisted
moduli. We have shown how the results may be applied to $\tilde{g}MSB$,
and discussed two explicit scenarios for this based
on two intersecting $5_1$ and $5_2$ brane sectors,
in which the MSSM gauge group is placed on the $5_2$ sector.
The second scenario in which $\tilde{g}MSB$ only applies to the
first two families, and the third family receives an unsuppressed
soft mass was first discussed in \cite{rayner}.

The general results will
be useful in phenomenological studies involving a combination of gravity and 
gaugino mediated SUSY breaking due to the dilaton, untwisted and twisted 
moduli contributions, and enable the soft masses to be studied
as a function of the compactification radii.
Previous analyses~\cite{allanach} have only considered
the effect of twisted moduli in 
the case where the gauge group and matter fields live on the
D9-brane, and share the same world-volume with all twisted moduli fields.  
However such a scenario does not give rise to localised matter fields 
(confined at intersection points) and in general one does not
encounter states which are sequestered from twisted moduli.
Hence the standard K\"{a}hler
potentials used in those analyses are perfectly acceptable. 
By contrast our analysis opens the door for more general type I string 
constructions involving $D9$ {\em and} $D5_i$-branes, 
where potentially more realistic phenomenology
and hierarchies between observables can be obtained with some or 
all of the matter fields sequestered 
from twisted moduli SUSY breaking sectors.

\acknowledgments
S.K. and D.R. would like to thank PPARC for a Senior Fellowship and a 
Studentship.  We would also like to thank the IPPP for their hospitality 
during the early stages of this work.  We thank Thomas Dent for
bringing our attention to ref.~\cite{altexp} and the non-perturbative
origin of the alternative suppression factor.

\appendix
\section{Supergravity Basics}  \label{sec:sugra}

We will now use a conventional supergravity formalism to describe the 4d
effective theory that arises as the low energy limit of the type I theory.
Supergravity (local SUSY) is defined in terms of a K\"{a}hler 
function (G) of generic chiral superfields $(\phi=h,C_{a})$ including the 
hidden sector closed strings $(h=S,T_{i},Y_{k})$ and open string matter states
 $(C_{a}=C^{5_{i}}_{j}, C^{5_{i} 5_{j}})$~\footnote{Notice that we have 
included powers of the reduced Planck mass $(\tilde{M}_{P})$ that appear in 
the K\"{a}hler function to obtain the correct dimensions, although it is 
conventional to adopt natural units and set $\tilde{M}_{P}=1$.}:
\begin{eqnarray}
 G(\phi, \bar{\phi}) = \frac{K(\phi,\bar{\phi})}{\tilde{M}_{P}^{2}} 
  + \ln \left(\frac{W(\phi)}{\tilde{M}_{P}^{3}} \right) + 
   \ln \left( \frac{W^{\ast}(\bar{\phi})}{\tilde{M}_{P}^{3}} \right)
 \label{eq:gkahler}
\end{eqnarray}
The K\"{a}hler potential $K(\phi,\bar{\phi})$ is a real function of 
chiral superfields and may be expanded in powers of matter states 
$C_{a}$~\cite{ibanez97} (including non-perturbative contributions):
\begin{eqnarray}
 K=\bar{K}(h,\bar{h}) +\tilde{K}_{\bar{a}b}(h,\bar{h}) \bar{C}_{\bar{a}} C_{b}
  + \left[ \frac{1}{2} Z_{ab}(h,\bar{h}) C_{a} C_{b} + h.c. \right] + \ldots
   \label{eq:kahexp}
\end{eqnarray}
where $\tilde{K}_{a\bar{b}}$ is the (generally non-diagonal) matter metric and
a non-zero bilinear term $Z_{ab}$ can generate the $\mu$-term 
through the Guidice-Masiero mechanism~\cite{gm} subject to gauge-invariance.
The superpotential $W(\phi)$ is a holomorphic function of chiral superfields 
that can also be expanded:
\begin{eqnarray}
 W=\hat{W}(h) + \frac{1}{2} \mu_{ab}(h) C_{a}C_{b} 
  + \frac{1}{6} Y_{abc} C_{a}C_{b}C_{c} + \ldots  \label{eq:supexp}
\end{eqnarray}
Notice that it includes a trilinear Yukawa term (that will generate fermion 
masses) and a bilinear $\mu$-term.  However, the K\"{a}hler potential and
superpotential also receive non-perturbative contributions that are often 
difficult to predict.  To make progress we will utilise a simple 
parametrisation of our ignorance of the non-perturbative sector in terms of 
Goldstino angles and CP-phases.

\subsection{Supergravity Potential}  \label{sec:sugrapot}

We know that in Nature SUSY must be broken and various mechanisms 
have been proposed.  It is convenient to analyse the SUSY breaking by 
considering the 
F-part of the SUGRA scalar potential~\footnote{We will ignore the D-term 
contribution to the potential that arises from the gauge sector.}.
It can be expressed in terms of derivatives of the K\"{a}hler function 
$G(\phi, \bar{\phi})$, or equivalently in terms of the F-term auxiliary fields
that can acquire non-zero vevs and trigger the SUSY breaking~\cite{ibanez97}.
Using Eq.\ref{eq:gkahler} we obtain:
\begin{eqnarray}
 V(\phi, \bar{\phi}) 
  = e^{G} \left[ G_{I} (K^{-1})_{I \bar{J}} G_{\bar{J}}
   - 3 \right] = F_{\bar{J}} \, K_{\bar{J} I} \, F_{I} 
    - 3 e^{K} \, |W|^{2}  \label{eq:potentialf}
\end{eqnarray}
where $I,J \equiv \phi_{I}, \phi_{J} \in S,T_{i},Y_{k},C_{a}$ and
\begin{eqnarray}
 G_{I} &\equiv& \frac{\partial G}{\partial \phi_{I}} 
  = \frac{W_{I}}{W} + K_{I} \\
 F_{I} &=& e^{G/2} (K^{-1})_{I \bar{J}} \, G_{\bar{J}}
\end{eqnarray}
where $(K^{-1})_{I\bar{J}}$ is the inverse of the metric $K_{\bar{J}I}$, and 
satisfies the relation
$(K^{-1})_{I \bar{J}} K_{\bar{J} L} = \delta_{IL}$.  A subscript on G denotes
partial differentiation, while the same subscript on F is just a label.  A 
barred subscript on an F-term denotes its conjugate field $F_{\bar{I}} \equiv 
\left( F_{I} \right)^{\dagger}$. We make no distinction between
upper and lower indices. 

After SUSY breaking, the supersymmetric partner of the 
Goldstone boson (Goldstino) is {\it eaten} by the massless gravitino through
the super-Higgs mechanism.  The gravitino now has a mass given by
\begin{eqnarray}
 m_{3/2}^{2} = e^{\langle G \rangle} = 
  e^{\langle K \rangle} \, |\langle W \rangle|^{2}
   = \frac{1}{3} \langle F_{\bar{J}} \, K_{\bar{J} I} \, F_{I} \rangle
\end{eqnarray}
and sets the overall scale of the soft parameters. 

In the absence of F-term vacuum expectation values ($\langle F_{I} \rangle = 
0 \,\, \forall \, \phi_{I}$), the locally supersymmetric vacuum is negative
$V_{SUSY}= - 3 e^{G}$.  However if one (or more) of the auxiliary F-terms
acquires a non-zero vev, the negative vacuum energy can be (partially) 
cancelled.  This raises the exciting possibility that the vacuum energy, or
rather the cosmological constant $V_{0}$, can be made vanishingly small in 
agreement with experimental limits.  Notice that such a possibility cannot 
arise in {\it global} SUSY.

\subsection{SUSY breaking F-terms}  \label{sec:ft}

As previously mentioned, (unknown) non-perturbative contributions to the
K\"{a}hler function require a parametrisation of our ignorance in terms of
Goldstino angles and CP-phases that {\it control} the relative contributions
to SUSY breaking from the various F-terms vevs.  We can define a column vector
of F-term vevs $F$ in terms of a matrix P and column vector $\Theta$ (which
also includes a CP-phase), where $\Theta$ has unit length and satisfies 
$\Theta^{\dagger} \Theta =1$, and P canonically normalises the
K\"{a}hler metric $P^{\dagger} K_{\bar{J} I} P = 1$:
\begin{eqnarray}
 F &=& \sqrt{3} \, C \, m_{3/2} \, \left( P \Theta \right)  \label{eq:genf} \\
 F^{\dagger} &=& \sqrt{3} \, C \, m_{3/2} \, 
  \left( \Theta^{\dagger} P^{\dagger} \right)  \nonumber
\end{eqnarray}

\noindent Replacing the fields by their vevs, we can rewrite 
Eq.\ref{eq:potentialf} as a matrix equation:
\begin{eqnarray}
 \langle V \rangle \equiv V_{0} &=& F^{\dagger} \, K_{\bar{J} I} \, F 
    - 3 m_{3/2}^{2}  \nonumber  \\
 &=& 3 C^{2} m_{3/2}^{2} \, \Theta^{\dagger} \Theta 
  \left( P^{\dagger} K_{\bar{J} I} P \right) - 3 m_{3/2}^{2} 
   \label{eq:potentialf2} \\
 &=& 3 m_{3/2}^{2} \left( C^{2} - 1 \right) \nonumber
\end{eqnarray}
where $V_{0}$ is the cosmological constant and hence
$C^{2}=1+\frac{V_{0}}{3m_{3/2}^{2}}$.  Therefore, choosing a vanishingly 
small cosmological constant sets $C=1$.

As an example consider the case of the dilaton $S$ and an overall moduli 
field $T$ with diagonal K\"{a}hler metric.  The SUGRA potential would be 
a ``sum of squares'' 
$V_{F} \sim |F_{S}|^{2} + |F_{T}|^{2} + \ldots -3e^{G}$
and hence the P-matrix is a diagonal normalising matrix: 
\begin{eqnarray}
 P_{I \bar{J}} = (K_{I \bar{I}})^{-1/2} \delta_{I \bar{J}}
\end{eqnarray}
In this special case we would recover the expressions of 
refs.\cite{ibanez97,ibanez94,ibanez98}:
\begin{eqnarray}
 F\equiv \left(
  \begin{array}{c}
   F_{S} \\
   F_{T} 
  \end{array} \right) = \sqrt{3}C \, m_{3/2} \left(
   \begin{array}{c}
    (K_{S \bar{S}})^{-1/2} \sin\theta \, e^{i\alpha_{S}} \\
    (K_{T \bar{T}})^{-1/2} \cos\theta \, e^{i\alpha_{T}}
   \end{array} \right)
\end{eqnarray}
so that dilaton(moduli) dominated SUSY breaking corresponds to $\sin\theta
(\cos\theta) =1$ respectively.
However in the more general case, the potential includes terms that mix 
different F-terms.  The action of the P-matrix is to canonically normalise
the K\"{a}hler metric and maintain the validity of the 
parametrisation~\footnote{The K\"{a}hler metric always receives off-diagonal
components from the matter fields, but these are conventionally assumed to
be small in comparison to the diagonal entries.  However, the anomaly
cancelling Green-Schwarz term mixes different fields at the same level to
introduce off-diagonal components of comparable size.}.

\subsection{Soft Masses and trilinears}  \label{sec:soft}

Using Eqs.\ref{eq:kahexp},\ref{eq:supexp} we can write down the 
{\it un-normalised} SUSY breaking masses and trilinears that arise in the 
soft SUGRA potential~\cite{ibanez97}:
\begin{eqnarray}
 V_{soft}=m^{2}_{\bar{a}b} \bar{C}_{\bar{a}} C_{b}
  + \left( \frac{1}{6} A_{abc} Y_{abc} C_{a}C_{b}C_{c} + h.c. \right) + \ldots
\end{eqnarray}
where the K\"{a}hler metrics are in general not diagonal leading to
the non-canonically normalised soft masses
\begin{eqnarray}
 m^{2}_{\bar{a}b} &=& \left( m_{3/2}^{2}+V_{0} \right) \tilde{K}_{\bar{a}b}
  - F_{\bar{m}} \left( \partial_{\bar{m}} \partial_{n} \tilde{K}_{\bar{a}b}
   - \partial_{\bar{m}} \tilde{K}_{\bar{a}c} (\tilde{K}^{-1})_{c\bar{d}}
    \partial_{n} \tilde{K}_{\bar{d}b} \right) F_{n}  \hspace*{1cm} \\
 A_{abc} Y_{abc} &=& \frac{\hat{W}^{\ast}}{|\hat{W}|} \, e^{\bar{K}/2} F_{m}
  \left[ \bar{K}_{m} Y_{abc} + \partial_{m} Y_{abc} 
   - \left( (\tilde{K}^{-1})_{d\bar{e}} \partial_{m} \tilde{K}_{\bar{e}a}
    Y_{dbc} \right. \right. \\
  & & \hspace*{65mm}\left.\left. + (a \leftrightarrow b) 
   + (a \leftrightarrow c) \right) \right]  \nonumber
\end{eqnarray}
where the subscript $m=h,C_{a}$.  
Notice that a non-diagonal K\"{a}hler metric
for the matter states will generate a mass matrix between different fields.
The physical masses and states are obtained by transforming to the
canonically normalised K\"{a}hler metric,
\begin{eqnarray}
 \tilde{K}_{\bar{a}b} \bar{C}_{\bar{a}} C_{b} \longrightarrow 
  \bar{C}'_{\bar{a}}C'_{a}.
\end{eqnarray}
The K\"{a}hler metric is canonically normalised by a transformation
$\tilde{P}^{\dagger} \tilde{K} \tilde{P} = 1$, so that
the physical canonically normalised
masses $m^{2}_{a}$ are related to the previous
non-canonical mass matrix $m^{2}_{\bar{a}b}$ by the relation
\begin{eqnarray}
 m^{2}_{a}
  =\tilde{P}^{\dagger} m^{2}_{\bar{a}b} \tilde{P}.
\end{eqnarray}
If the K\"{a}hler matter metric is diagonal (but not canonical)
$\tilde{K}_{a}=\tilde{K}_{\bar{a}b}\delta_{\bar{a}b}$
then the canonically normalised scalar masses $m_{a}^{2}$ are
simply given by
\begin{eqnarray}
 m_{a}^{2}= m_{3/2}^{2} 
  - F_{\bar{J}} F_{I} \partial_{\bar{J}} \partial_{I} 
   \left( \ln \tilde{K}_{a} \right)   \hspace*{1cm} (I,J=h,C_{a}). 
    \label{eq:scmass}
\end{eqnarray}
The soft gaugino mass associated with the gauge group $G_{\alpha}$ is:
\begin{eqnarray}
 M_{\alpha} = \frac{1}{2 \, Re f_{\alpha}} F_{I} \partial_{I} f_{\alpha}
  \hspace*{1cm} (I=S,T_{i}, Y_{k})  \label{eq:gauginomass2}
\end{eqnarray}
and the canonically normalised
SUSY breaking trilinear term for the scalar fields\\ 
\noindent $A_{abc} Y_{abc} C_{a} C_{b} C_{c}$ is
\begin{eqnarray}
 A_{abc} = F_{I} \left[ \overline{K}_{I} + \partial_{I} \ln Y_{abc}
  - \partial_{I} \ln \left( \tilde{K}_{a} \tilde{K}_{b} \tilde{K}_{c} \right)
   \right].
    \label{eq:trilinear2}
\end{eqnarray}

\section{Alternative exponential suppression factor}  \label{sec:alt}

In this appendix we consider an alternative suppression factor 
$e^{-\left(M_{\ast}r\right)^{2}}$ that is attributed to
non-perturbative world-sheet instanton corrections~\cite{altexp}, and
differs from the (previous) field theory interpretation of the
suppression due to propagating massive modes.

We summarize the modified scalar masses and trilinears found by
repeating the earlier calculations in section \ref{sec:softp}, but with
the alternative suppression factor.  The non-sequestered scalar masses
of Eq.~\ref{eq:52mass} remain unchanged, but Eq.~\ref{eq:51mass} becomes:
\begin{eqnarray} 
 m_{C^{5_{1} 5_{2}}}^{2} &=& \tilde{m}^{2} - \frac{3}{2} m_{3/2}^{2} 
  \left( \sin^{2} \theta + \Theta_{3}^{2} \, \cos^{2} \theta \, \sin^{2} \phi 
   \right)  \nonumber \\
 m_{C^{9 5_{1}}}^{2} &=& \tilde{m}^{2} - \frac{3}{2} m_{3/2}^{2} \,
  \cos^{2}\theta \, \sin^{2}\phi \left( \frac{\Theta_{2}^{2}}{k}
   + \Theta_{3}^{2} \right) \nonumber \\
m_{C^{5_{1}}_{1}}^{2} &=& \tilde{m}^{2} - 3 m_{3/2}^{2} \sin^{2}\theta 
   \label{eq:51massapp} \\
 m_{C^{5_{1}}_{2}}^{2} &=& \tilde{m}^{2} - 3 m_{3/2}^{2} \Theta_{3}^{2} 
  \cos^{2}\theta \sin^{2}\phi  \nonumber \\
 m_{C^{5_{1}}_{3}}^{2} &=& \tilde{m}^{2} - \frac{3}{k} m_{3/2}^{2} 
  \Theta_{2}^{2} \cos^{2}\theta \sin^{2}\phi  \nonumber
\end{eqnarray}
where  
\begin{eqnarray}
 \tilde{m}^{2} = m_{3/2}^{2} \left[ \vrule width 0pt height 18pt
  1 - \cos^{2}\theta \cos^{2}\phi
    \left( 1-e^{-(T_{2}+\bar{T}_{2})/4} \right) \right. \hspace*{7cm} 
     \nonumber \\
 - \frac{\cos^{2}\theta \sin^{2}\phi \, \Theta_{2}^{2} \, \delta_{GS}}{k}
  \left( 1- e^{-(T_{2}+\bar{T}_{2})/4} \right)
   \left\{ Y_{2}+\bar{Y_{2}}-\delta_{GS} \ln(T_{2}+\bar{T}_{2}) \right\} 
    \hspace*{1.2cm}  \label{mtildeapp} \\
 + \frac{\cos^{2}\theta \sin^{2}\phi \, \Theta_{2}^{2} \, 
  e^{-(T_{2}+\bar{T}_{2})/4}
  }{32 k} (T_{2}+\bar{T}_{2})^{2} \left\{ Y_{2}+\bar{Y_{2}}-\delta_{GS} 
   \ln(T_{2}+\bar{T}_{2}) \right\}^{2} \hspace*{0.8cm} \nonumber \\
 - \frac{\cos^{2}\theta \cos\phi \sin\phi \, \left( \Theta_{2} \, 
  e^{i(\alpha_{2}-\alpha_{Y_{2}})} + \Theta_{2}^{\dagger} \, 
   e^{-i(\alpha_{2}-\alpha_{Y_{2}})} \right) e^{-(T_{2}+\bar{T}_{2})/4} 
    }{32 \sqrt{k} }  \hspace*{0.4cm} \nonumber \\
 \times \left\{ Y_{2}+\bar{Y_{2}}-\delta_{GS} \ln(T_{2}+\bar{T}_{2}) \right\} 
  \left. \left( \vrule width 0pt height 15pt
  8 (T_{2}+\bar{T}_{2}) + \delta_{GS} 
   \left\{ Y_{2}+\bar{Y_{2}}-\delta_{GS} \ln(T_{2}+\bar{T}_{2}) \right\} 
    \vrule width 0pt height 16pt\right)
     \vrule width 0pt height 18pt \right] \nonumber
\end{eqnarray}
which replaces Eq.~\ref{mtilde}; and the masses are expanded up to 
${\mathcal O}\left[ \frac{\delta_{GS}}{T_{2}+\bar{T}_{2}} \right]$.

Similarly we can find modified expressions for the trilinears of
Eqs.~\ref{eq:aii23} - \ref{eq:a9123}. 
\begin{eqnarray}
 A_{C^{5_{2}}_{3} C^{5_{1}5_{2}} C^{5_{1}5_{2}}} 
  = -\sqrt{3} m_{3/2} \, \cos\theta \left[ \sin\phi \,
  \frac{\Theta_{2} \, e^{i\alpha_{2}}}{\sqrt{k}}  \right. \hspace*{6.6cm}
   \nonumber \\
 + \sin\phi \, \frac{\Theta_{2} \, e^{i\alpha_{2}}}{12\sqrt{k}} \,
  e^{-(T_{2}+\bar{T}_{2})/4} \, (T_{2}+\bar{T}_{2})
   \left\{ Y_{2}+\bar{Y_{2}}-\delta_{GS} 
   \ln(T_{2}+\bar{T}_{2}) \right\}^{2}  \label{eq:aii23app} \hspace*{1.5cm} \\
 \left. - \cos\phi \, \frac{e^{i\alpha_{Y_{2}}}}{3}
  \left( 1+ 2 \, e^{-(T_{2}+\bar{T}_{2})/4} \right) 
   \left\{ Y_{2}+\bar{Y_{2}}-\delta_{GS} \ln(T_{2}+\bar{T}_{2}) \right\} 
\right] \hspace*{1cm}
  \nonumber \\
 A_{C^{9}_{1} C^{9 5_{1}} C^{9 5_{1}}} 
  = -\sqrt{3} m_{3/2} \, \left[ \sin\theta \, e^{i\alpha_{S}}
   \right. \hspace*{8.7cm} \nonumber \\
 + \cos\theta \sin\phi \, \frac{\Theta_{2} \, e^{i\alpha_{2}}}{12\sqrt{k}} \,
  e^{-(T_{2}+\bar{T}_{2})/4} \, (T_{2}+\bar{T}_{2})
   \left\{ Y_{2}+\bar{Y_{2}}-\delta_{GS} 
   \ln(T_{2}+\bar{T}_{2}) \right\}^{2}  \label{eq:a9951951app} 
    \hspace*{1.5cm} \\
 \left. - \cos\theta \cos\phi \, \frac{e^{i\alpha_{Y_{2}}}}{3}
  \left( 1+ 2 \, e^{-(T_{2}+\bar{T}_{2})/4} \right) 
   \left\{ Y_{2}+\bar{Y_{2}}-\delta_{GS} \ln(T_{2}+\bar{T}_{2}) \right\} 
    \right] \hspace*{1cm} \nonumber \\
 A_{C^{5_{1} 5_{2}} C^{9 5_{1}} C^{9 5_{2}}} 
  = -\sqrt{3} m_{3/2} \, \left[ \frac{1}{2} \sin\theta \, e^{i\alpha_{S}}
   + \frac{1}{2} \cos\theta\sin\phi \left( \Theta_{1} \, e^{i\alpha_{1}} +
  \frac{\Theta_{2} \, e^{i\alpha_{2}}}{\sqrt{k}} 
   - \Theta_{3} \, e^{i\alpha_{3}} \right) \right. \hspace*{2mm} \nonumber \\
 + \cos\theta \sin\phi \, \frac{\Theta_{2} \, e^{i\alpha_{2}}}{12\sqrt{k}} \,
  e^{-(T_{2}+\bar{T}_{2})/4} \, (T_{2}+\bar{T}_{2})
   \left\{ Y_{2}+\bar{Y_{2}}-\delta_{GS} 
   \ln(T_{2}+\bar{T}_{2}) \right\}^{2}  \label{eq:ai951952app} 
    \hspace*{1cm} \\
 \left. - \cos\theta \cos\phi \, \frac{e^{i\alpha_{Y_{2}}}}{3}
  \left( 1+ 2 \, e^{-(T_{2}+\bar{T}_{2})/4} \right) 
   \left\{ Y_{2}+\bar{Y_{2}}-\delta_{GS} \ln(T_{2}+\bar{T}_{2}) \right\} 
    \right] \hspace*{5mm} \nonumber \\
 A_{C^{5_{1}}_{3} C^{5_{1}5_{2}} C^{5_{1}5_{2}}} = 
  A_{C^{5_{1}}_{1} C^{5_{1}}_{2} C^{5_{1}}_{3}} = 
   A_{C^{5_{1}}_{1} C^{9 5_{1}} C^{9 5_{1}}} 
    = -\sqrt{3} m_{3/2} \, \cos\theta \left[ \sin\phi \, \Theta_{1} \, 
     e^{i\alpha_{1}} \right. \nonumber \hspace*{1.8cm} \\
 + \sin\phi \frac{\Theta_{2} \, e^{i\alpha_{2}}}{8\sqrt{k}} \, 
  e^{-(T_{2}+\bar{T}_{2})/4} \, (T_{2}+\bar{T}_{2})
   \left\{ Y_{2}+\bar{Y_{2}}-\delta_{GS} 
    \ln(T_{2}+\bar{T}_{2}) \right\}^{2}  \label{eq:aii13app} 
     \hspace*{1.5cm} \\
 \left. - \cos\phi \, e^{i\alpha_{Y_{2}}}
  \, e^{-(T_{2}+\bar{T}_{2})/4} 
   \left\{ Y_{2}+\bar{Y_{2}}-\delta_{GS} \ln(T_{2}+\bar{T}_{2}) \right\} 
    \right] \hspace*{1cm} \nonumber \\
 A_{C^{5_{2}}_{1} C^{5_{2}}_{2} C^{5_{2}}_{3}} 
  = A_{C^{5_{2}}_{2} C^{9 5_{2}} C^{9 5_{2}}} 
   = -\sqrt{3} m_{3/2} \, \cos\theta \left[ \sin\phi \,
   \frac{\Theta_{2} \, e^{i\alpha_{2}}}{\sqrt{k}} \right. \hspace{4.6cm} 
    \nonumber \\
 \left. - \cos\phi \, e^{i\alpha_{Y_{2}}}
  \left\{ Y_{2}+\bar{Y_{2}}-\delta_{GS} \ln(T_{2}+\bar{T}_{2}) \right\} 
   \right] \label{eq:a212223app} \hspace*{1.5cm} \\
 A_{C^{9}_{1} C^{9}_{2} C^{9}_{3}} = A_{C^{9}_{2} C^{9 5_{2}} C^{9 5_{2}}} 
  = -\sqrt{3} m_{3/2} \, \left[ \sin\theta \, e^{i\alpha_{S}} \right.
   \hspace*{6.7cm} \nonumber \\
 \left. - \cos\theta\cos\phi \, e^{i\alpha_{Y_{2}}}
  \left\{ Y_{2}+\bar{Y_{2}}-\delta_{GS} \ln(T_{2}+\bar{T}_{2}) \right\} 
   \right] \label{eq:a9123app} \hspace*{1.5cm}
\end{eqnarray}
where all trilinears have been expanded up to ${\mathcal O} \left( 
\frac{1}{T_{2}+\bar{T}_{2}} \right)$.

It is now straightforward to consider these modified expressions in
different limits of SUSY breaking domination as before.



\begin{thebibliography}{99}
 \bibitem{dual} A. Font, L.E. Ib\'{a}\~{n}ez, D. L\"{u}st and F. Quevedo,
                 {\it Phys. Lett.} {\bf B249} (1990) 35; M. Duff and J. Lu, 
                 {\it Nucl. Phys.} {\bf B357} (1991) 534; A. Sen, 
                 {\it Int. J. Mod. Phys.} {\bf A9} (1994) 3707, 
                 [hep-th/9402002].
 \bibitem{dbranes} For a review, see: J. Polchinski, hep-th/9611050.
 \bibitem{ibanez97} A. Brignole, L.E. Ib\'{a}\~{n}ez and C. Mu\~{n}oz,
                    hep-ph/9707209.
 \bibitem{ibnil} L.E. Ib\'{a}\~{n}ez and H.P. Nilles, {\it Phys. Lett.}
                  {\bf B169} (1986) 354.
 \bibitem{stronghet} T. Banks and M. Dine, {\it Nucl. Phys.} {\bf B479} (1996)
                      173, [hep-th/9605136]; K. Choi, {\it Phys. Rev.}
                      {\bf D56} (1997) 6588, [hep-th/9706171]; H.P. Nilles
                      and S. Stieberger, {\it Nucl. Phys.} {\bf B499} (1997)
                      3, [hep-th/9702110]; H.P. Nilles, M. Olechowski and
                      M. Yamaguchi, {\it Phys. Lett.} {\bf B415} (1997) 24,
                      [hep-th/9707143].
 \bibitem{gs} L.E. Ib\'{a}\~{n}ez, R. Rabadan and A.M. Uranga,
               {\it Nucl. Phys.} {\bf B542} (1999) 112, [hep-th/9808139].
 \bibitem{ibanez94} A. Brignole, L.E. Ib\'{a}\~{n}ez and C. Mu\~{n}oz,
                    {\it Nucl. Phys.} {\bf B422} (1994) 125, [hep-ph/9308271];
                     Erratum-ibid. {\bf B436} (1995) 747; A. Brignole,
                     L.E. Ib\'{a}\~{n}ez, C. Munoz and C. Scheich, 
                     {\it Z. Phys.} {\bf C74} (1997) 157, [hep-ph/9508258].
 \bibitem{ibanez98} L.E. Ib\'{a}\~{n}ez, C. Mu\~{n}oz and S. Rigolin, 
                     {\it Nucl. Phys.} {\bf B553} (1999) 43, [hep-ph/9812397].
 \bibitem{allanach} S.A. Abel, B.C. Allanach, L.E. Ib\'{a}\~{n}ez, M. Klein 
                     and F. Quevedo, {\it JHEP} {\bf 0012} (2000) 026,
                     [hep-ph/0005260]; B.C. Allanach, D. Grellscheid and 
                     F. Quevedo, hep-ph/0111057.
 \bibitem{benakli} K. Benakli, {\it Phys. Lett.} {\bf B475} (2000) 77,
                    [hep-ph/9911517].
 \bibitem{gaugino} D.E. Kaplan, G. Kribs and M. Schmaltz, {\it Phys. Rev.} 
                    {\bf D62} (2000) 035010, [hep-ph/9911293]; Z. Chacko, 
                    M. Luty, A.E. Nelson and E. Pont\'{o}n, 
                    {\it JHEP} {\bf 0001} (2000) 003, [hep-ph/9911323].
 \bibitem{rs} L.Randall and R.Sundrum, {\it Nucl. Phys.} {\bf B557} (1999)
               79, [hep-th/9810155].
 \bibitem{simple} E. Poppitz, {\it Nucl. Phys.} {\bf B542} (1999) 31, 
                   [hep-th/9810010]; C.A. Scrucca and M. Serone,
                   {\it JHEP} {\bf 0007} (2000) 025, [hep-th/0006201].
 \bibitem{abel} S.A. Abel and G. Servant, {\it Nucl. Phys.}
                  {\bf B597} (2001) 3, [hep-th/0009089]; S.A. Abel and 
                  G. Servant, {\it Nucl. Phys.} {\bf B611} (2001) 43, 
                  [hep-ph/0105262].
 \bibitem{altexp} S. Hamidi and C. Vafa, {\it Nucl. Phys.} {\bf B279}
                   (1987) 465; L.J. Dixon et al., {\it Nucl. Phys.}
                   {\bf B282} (1987) 13; L.E. Ib\'{a}\~{n}ez, {\it
                   Phys. Lett.} {\bf B181} (1986) 269. 
 \bibitem{gordy} L.Everett, G.L.Kane and S.F.King, {\it JHEP}
                  {\bf 008} (2000) 012, [hep-ph/0005204].
 \bibitem{shiutye} G. Shiu and S.H.H Tye, {\it Phys. Rev.} {\bf D58} 
                    (1998) 106007, [hep-th/9805157].
 \bibitem{rayner} S.F. King and D.A.J. Rayner, {\it Nucl. 
                   Phys.} {\bf B607} (2001) 77, [hep-ph/0012076].
 \bibitem{gm} G.F. Giudice and A. Masiero, {\it Phys. Lett.} {\bf B206} (1988)
               480.
\end{thebibliography}
\end{document}